\documentclass[reprint,twocolumn,notitlepage,prl,superscriptaddress,showpacs]{revtex4-1}
\usepackage{hyperref}
\usepackage[usenames,dvipsnames]{color}
\usepackage{amsmath}
\usepackage[english]{babel}
\usepackage{amssymb}
\usepackage{graphicx}
\usepackage{epstopdf}
\usepackage{comment}
\usepackage{mathrsfs}
\usepackage{amsfonts}
\usepackage{tabularx}
\usepackage{multirow}
\usepackage[]{todonotes}
\usepackage[]{algorithm2e}

\begin{document}

\title{Algorithmic approach to diagrammatic expansions for real-frequency evaluation of susceptibility functions}
\author{Amir Taheridehkordi}
\affiliation{Department of Physics and Physical Oceanography, Memorial University of Newfoundland, St. John's, Newfoundland \& Labrador, Canada A1B 3X7} 
\author{S. H. Curnoe}
\affiliation{Department of Physics and Physical Oceanography, Memorial University of Newfoundland, St. John's, Newfoundland \& Labrador, Canada A1B 3X7}
\author{J. P. F. LeBlanc}
\email{jleblanc@mun.ca}
\affiliation{Department of Physics and Physical Oceanography, Memorial University of Newfoundland, St. John's, Newfoundland \& Labrador, Canada A1B 3X7} 

\date{\today}
\begin{abstract}
 We systematically generate the perturbative expansion for the two-particle spin susceptibility in the Feynman diagrammatic formalism and apply this expansion to a model system - the single-band Hubbard model on a square lattice. We make use of algorithmic Matsubara integration (AMI)  [A. Taheridehkordi, S. H. Curnoe, and J. P. F. LeBlanc, Phys. Rev. \textbf{B} 99, 035120 (2019)] to analytically evaluate  Matsubara frequency summations, allowing us to symbolically impose analytic continuation to the real frequency axis. We minimize our computational expense by applying graph invariant transformations [Amir Taheridehkordi, S. H. Curnoe, and J. P. F. LeBlanc,
Phys. Rev. \textbf{B} 101, 125109 (2020)].  We highlight extensions of the random-phase approximation and T-matrix methods that, due to AMI, become tractable.
We present results for weak interaction strength where the direct perturbative expansion is convergent, and verify our results on the Matsubara axis by comparison to other numerical methods. By examining the spin susceptibility as a function of real-frequency via an order-by-order expansion we can identify precisely what role higher order corrections play on spin susceptibility and demonstrate the utility and limitations of our approach.
\end{abstract}

\maketitle

The Hubbard model \cite{Hubbard1963} has become a laboratory for  the development of numerical tools in correlated electron systems. The single-band model on a two-dimensional (2D) square lattice is believed to be the minimal model to capture features of high-temperature superconductivity \cite{MANCINI1995361} yet remains a complex numerical problem that has motivated the development of numerous novel numerical algorithms \cite{benchmarks,Schaefer:2020}.

The single-particle properties of that model have been investigated by a wide variety of different methods, from non-perturbative approaches such as dynamical Mean-field theory (DMFT) \cite{Kotliar01,Kotliar:2006} and dynamical cluster approximation (DCA) \cite{hettler:1998,hettler:2000} to perturbative methods such as diagrammatic Monte Carlo (DiagMC) \cite{Prokof'ev:1998,vanhoucke,vanhoucke:natphys,kozik:2010,rossi2017determinant,Rossi:shiftedaction,ferrero:2018,Evgeny:2019}.
Understanding the role of two-particle excitations - for experiments on cuprates \cite{Takigawa1994,coldea:2001,Fujita:2012,Suzuki2018,Greco2019} as well as for numerical calculations of model systems \cite{fedor:2020,Chen:1994,Bulut,Macridin:2006,gunnarsson:2015,Chen:2015,Qin:2017,LeBlanc:2019,Hille:2020} - is of particular importance due to the subtle connections between spin excitations, antiferromagnetic order, superconductivity and pseudogap phenomena.
Despite the wide range of existing numerical algorithms the ability for numerical work to make concrete connections to experiment has been largely hampered by the challenges associated with evaluating the necessary two-particle spin and charge response functions.

There exists a greater issue that, in addition to the complexity of two
particle response functions, many numerical methods are constructed
around the finite-temperature Matsubara formalism and provide results in
an abstract imaginary time/Matsubara frequency space.  While results for
physically relevant properties on the so-called `real-time/frequency'
axis can be obtained but require numerical analytic
continuation procedures for which  solutions are not unique \cite{bergeron:2016,Levy2016,Gaenko17,ALPSCore,alpscore_v2}. As a result, the numerical analytic continuation process dominates the uncertainty of the result and compromises any attempt at high-precision numerics
\cite{Mravlje,Huang2019,Ferrero:AMI}.  In principle, this issue can be avoided through a
textbook application of the residue theorem to resolve the Matsubara summations,
resulting in analytical expressions for which analytic continuation is
the simple substitution of $i\Omega_n \rightarrow \Omega +i0^+$.
For low order diagrams this can be done by hand but for higher order corrections the resulting expressions become incomprehensibly complicated. For that reason this known solution is discarded for all but the most weakly correlated electron systems.
We have recently overcome this particular road-block with the method of
algorithmic Matsubara integration (AMI) \cite{AMI}, a procedure that
automates the construction of such analytic results and in principle allows
for a direct evaluation of arbitrary diagrammatic expansions composed of
thousands of analytic terms on the real-frequency axis.
In order to compute two-particle susceptibilities the number of diagrams to be evaluated using AMI is quite large and grows quickly with expansion order. In addition, there still remains a
general sign problem \cite{Loh,Chandrasekharan} as well as a more
fundamental fermionic sign due to cancellation between diagrams in the
expansion. In order to suppress the second issue, one opportunity lies
in the construction of \emph{sign-blessed} diagram groups by application
of graph invariant transformations (GIT) that can effectively be
combined with AMI \cite{GIT}.

In anticipation of these developments, we present the spin susceptibility
of the 2D Hubbard model,  \emph{beyond} random phase approximation (RPA) \cite{Bohm_RPA_1,Bohm_RPA_2}, T-matrix approximation (TMA) \cite{Fukuyama:1990,Gukelberger:2015} and low-order vertex corrections \cite{Yoshimi:2009}, in the real-frequency domain without need for any ill-posed numerical analytic continuation procedures.

\indent \emph{Hubbard model:}
We consider the single-band Hubbard Hamiltonian \cite{Lieb:1968,benchmarks},
\begin{eqnarray}\label{E:Hubbard}
H = \sum_{\langle ij\rangle \sigma} t_{ij}c_{i\sigma}^\dagger c_{j\sigma} + U\sum_{i} n_{i\uparrow} n_{i\downarrow} -\mu\sum_{i\sigma}n_{i\sigma},
\end{eqnarray}
where $t_{ij}$ is the hopping amplitude, $c_{i\sigma}^{(\dagger)}$ is the annihilation (creation) operator at site $i$, $\sigma \in \{\uparrow,\downarrow\}$ is the spin, $U$ is the onsite Hubbard interaction, $n_{i\sigma} = c_{i\sigma}^{\dagger}c_{i\sigma}$ is the number operator,  $\mu$ is the chemical potential, and $\langle ij \rangle$ restricts the sum to nearest neighbors. For a two dimensional square lattice we take $t_{ij}=-t$, resulting in the free particle energy dispersion
\begin{eqnarray}
\epsilon(\textbf k)=-2t[\cos(k_x)+\cos(k_y)]-\mu.
\end{eqnarray}

\indent \emph{Transverse spin susceptibility:}
The expansion for the transverse spin susceptibility is straightforwardly represented in position (\textbf r) and imaginary time ($\tau$) space and is defined as \cite{Bulut}
\begin{eqnarray}\label{E:chiT_def}
\chi_T(x,x') = \langle \mathcal T S_+(x) S_-(x') \rangle,
\end{eqnarray}
where $\mathcal T$ is the time-ordering operator, $x=(\textbf r, \tau)$, and $S_{+/-}$ are spin-ladder operators which are given by
$S_+(x)=S_-^{\dagger}(x)=c^\dagger_\uparrow(x)c_{\downarrow}(x)$.
One could instead construct the diagrammatic series for the longitudinal spin susceptibility $\chi_L(x,x')=\langle \mathcal T S_z(x) S_z(x') \rangle$ \cite{Hille:2020}, however, the spin-rotation invariance of the Hubbard Hamiltonian \cite{Alvarez:1996,Masumizu:2005} implies that $\chi_T=2\chi_L$; we note that the diagrammatic series for the transverse spin susceptibility is substantially simpler.  

\indent \emph {Constructing diagrams and integrands:}
We use perturbation theory to evaluate the
transverse spin susceptibility defined by Eq.~(\ref{E:chiT_def}). 
We construct the perturbative expansion and using standard Wick decomposition we represent the result as a series of Feynman diagrams \cite{Feynman,Baym,Luttinger:1960} that can then be evaluated in the momentum and frequency space.
Each transverse susceptibility diagram in the series has the property that the particle lines in the \emph {principle loop} (a unique fermionic loop that involves the two external vertices) have spin $\uparrow$, while anti-particle lines have spin $\downarrow$. Furthermore, since the on-site Hubbard interaction only occurs between solid lines with different spins we only consider diagrams that satisfy this criterion. 

First, we systematically generate all the \emph{topologically distinct} transverse susceptibility diagrams up to a truncation order $m_c$ by following the procedure described in Ref.~\onlinecite{GIT}. In order to reduce the diagrammatic space we neglect all diagrams with tadpole insertions by applying the chemical potential shift $\mu \to \mu-\bar n U/2$, where $\bar n$ is the number of electrons per site \cite{Zlatic,Daul1997}.  
Generally, a  $m$th order susceptibility diagram will have $m$
interaction
lines and $2m+2$ fermionic lines; with fixed external frequency
$i\Omega$
and momentum $\textbf q$, energy and momentum conservation at each interaction
imply that
there will be $m+1$ independent (internal) frequencies and momenta.
Following the
method outlined in Refs.\ \onlinecite{AMI,GIT}, we assign frequency
($X^j$) and momenta ($\textbf K_j$) variables to each fermionic line,
where $X^j$ and $\textbf K_j$ are linear combinations of the
independent frequencies and momenta.
Applying the Feynman rules, a diagram $D_{\zeta_m}$ of order $m$ with
topology $\zeta_m$ is evaluated as:
\begin{align}
D_{\zeta_m}(i\Omega,\textbf q,\beta,\mu) = \frac{(-1)^{m+F_{\zeta_m}}U^m}{(2\pi)^{2m+2}\beta^{m+1}} \nonumber \\  \times\sum\limits_{\{\textbf k_{m+1}\}}  \sum \limits _{\{\nu_{m+1}\}}  \prod
\limits_{j=1}^{2m+2}   \mathcal G_0^j(\epsilon ^j, X^j ). \label{eqn:each_diag_in_goal}
\end{align}
Here, $F_{\zeta_m}$ is the number of fermionic loops, $\beta$ is the inverse temperature, $\{\textbf k_{m+1}\}$ and $\{\nu_{m+1}\}$ are sets of (independent) internal momenta and frequencies, respectively, $\epsilon^j=\epsilon(\textbf K_j)$ represents the particle dispersion energy of the $j$th line 
and $\mathcal G_0^j(\epsilon ^j, X^j )=(X^j-\epsilon^j)^{-1}$ is the bare Green's function assigned to the $j$th solid line.

We first symbolically evaluate the Matsubara sums in
Eq.~(\ref{eqn:each_diag_in_goal}) by utilizing the residue theorem.
Although conceptually straightforward, the complexity of the
resulting analytic
expressions requires an
automated machinery. For this we follow the AMI procedure, described in
Refs.~\onlinecite{AMI,GIT}, to automatically construct and store the
analytic expressions for the Matsubara sums.
\begin{figure}
\centering
\includegraphics[width=0.51\linewidth]{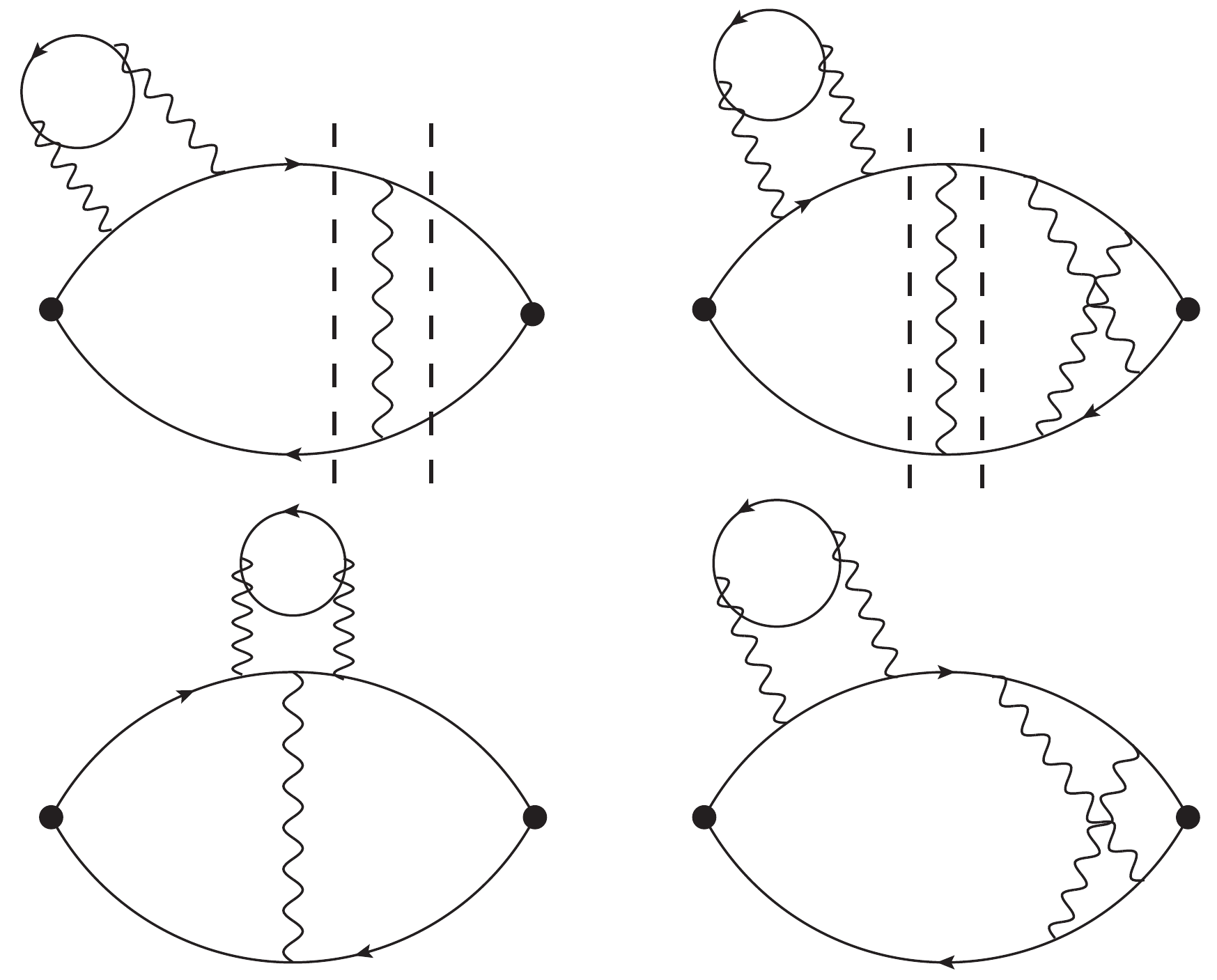}
\caption{\emph{Top row}: Examples of ladder-like diagrams which must be excluded from the ETM series. \emph{Bottom row}: Examples of non-ladder-like diagrams which must be included in the ETM series. 
The dashed lines identify where the ladder-like diagrams split into independent parts (there
are no common independent frequency-momenta variables to the left and
right). Solid and wavy lines are fermionic and interaction lines,
respectively.\label{fig:LL_NLL}}
\end{figure}
Finally, we use an integration procedure \cite{GIT,Zhiqiang} to evaluate the momenta sums of the diagrams.  Since the integrands are functions of continuous variables both Monte Carlo and deterministic approaches can be applied \cite{CUBA}, and so long as the internal $\{\textbf k_{m+1}\}$ space is not discretized the results are automatically in the thermodynamic limit. 

As an order by order expansion, the perturbative series of the transverse spin susceptibility is then written as 
\begin{align}
\chi_T^{(m_c)}(i\Omega,\textbf q,\beta,\mu)=  \sum\limits_{m=0}^{m_c}\sum\limits_{\zeta_m} D_{\zeta_m}(i\Omega,\textbf q,\beta,\mu), \label{eqn:goal}
\end{align}
where the sum over $\zeta_m$ is over all unique topologies of order $m$, here summed to a cutoff order $m_c$. 
The direct evaluation of Eq.~(\ref{eqn:goal}) is a challenging task due to the factorial increase of the number of diagrams with order \cite{Kugler} as well as a factorial increase in the number of integrated terms after applying AMI.  Therefore, to reduce the diagrammatic space we propose an alternative procedure, which we call \emph{extended T-matrix} (ETM), to approximate the transverse spin susceptibility by only evaluating a subset of susceptibility diagrams.  We categorize diagrams into two types: \emph{ladder-like} diagrams, those that can be factored into two (or more) independent integrals, and \emph{non-ladder-like} diagrams, which cannot be factored. Examples of such diagrams are shown in Fig.~\ref{fig:LL_NLL}.      
We define $\chi^{(m_c)}_{NL}$ to be the sum of all the non-ladder-like diagrams up to a truncation order $m_c$; then the transverse spin susceptibility is approximated by
\begin{eqnarray}\label{E:chiT_from_chi0}
\chi_{ETM}^{(m_c)}(i\Omega,\textbf q,\beta,\mu) = \frac{\chi^{(m_c)}_{NL}(i\Omega,\textbf q,\beta,\mu)}{1-U\chi^{(m_c)}_{NL}(i\Omega,\textbf q,\beta,\mu)}.
\end{eqnarray}
In a general sense, $\chi^{(m_c)}_{NL}$ and $U$ play the same roles in the transverse spin susceptibility expansion as the \emph{bare} Green's function and self-energy do in the diagrammatic expansion of the full Green's function \cite{tarantino:2017,Fetter}, but here for a very specific set of diagrams. 
Eq.~(\ref{E:chiT_from_chi0}) reduces to the RPA for longitudinal spin susceptibility and to the TMA for transverse spin susceptibility at $m_c=0$ ($\chi_{NL}^{(0)}$ is the bare bubble), while in the $m_c \to \infty$ limit it recovers the direct expansion of Eq.~(\ref{eqn:goal}). This provides a systematic bridge between those coarse approximations and the exact result and we  expect that Eq.~(\ref{E:chiT_from_chi0}) with $m_c \ge 1$ will provide more reliable results when compared to the RPA and TMA approaches. We therefore have two methods available to us: the direct order-by-order evaluation of all topologies via Eq.~(\ref{eqn:goal}) and the ETM scheme which results in fewer diagrams to be evaluated.
\begin{table}
\caption{Diagrammatic space reduction of the transverse spin susceptibility up to
fourth order at half-filling. In the second row, $n^{(m)}$ is the number of
diagrams at each order $m$ (not including diagrams with tadpole insertions), and $(n_{NL}^{(m)})$ is the number of non-ladder-like diagrams at each order $m$.
In the last row, $n_{g}^{(m)}$ is the number of groups of equal diagrams at each order $m$, and ($n_{g,NL}^{(m)})$ is the number of groups of equal non-ladder-like diagrams at each order $m$. }
\label{tab:GIT_Reduction}      %
\begin{center}
\begin{tabular}{|c|c|c|c|c|c|}
\hline
$m$ & 0 & 1 & 2 & 3 & 4 \\
\hline \hline
$n^{(m)}(n_{NL}^{(m)})$ & 1(1) & 1(0) & 4(3) & 17(10) & 101(22)  \\
\hline
$n_{g}^{(m)}(n_{g,NL}^{(m)})$ & 1(1) & 1(0) & 3(2) & 6(3) & 71(16) \\
\hline
\end{tabular}\\
\end{center}
\end{table}

By taking advantage of the
inherent symmetry of the half-filled Hubbard model on a square lattice
the diagrammatic space can be further reduced.
We apply the GIT procedure \cite{GIT} to identify \emph{exactly}
canceling
and \emph{exactly} equal diagrams at half-filling.  The
complete diagrammatic space reduction is shown in
Table~\ref{tab:GIT_Reduction}.
In order to calculate the transverse spin
susceptibility up to third order via Eq.~(\ref{eqn:each_diag_in_goal}),
we
need to evaluate only 11 diagrams in total. For the ETM approach this
number is further reduced to only six diagrams at third
order, and by 4th
order the number of non-ladder-like diagrams drops drastically with
$\approx 80\%$ of the diagrams being ladder-like.
\indent \emph{Numerical results and comparisons:}
\begin{figure}
\centering
\includegraphics[width=0.85\linewidth]{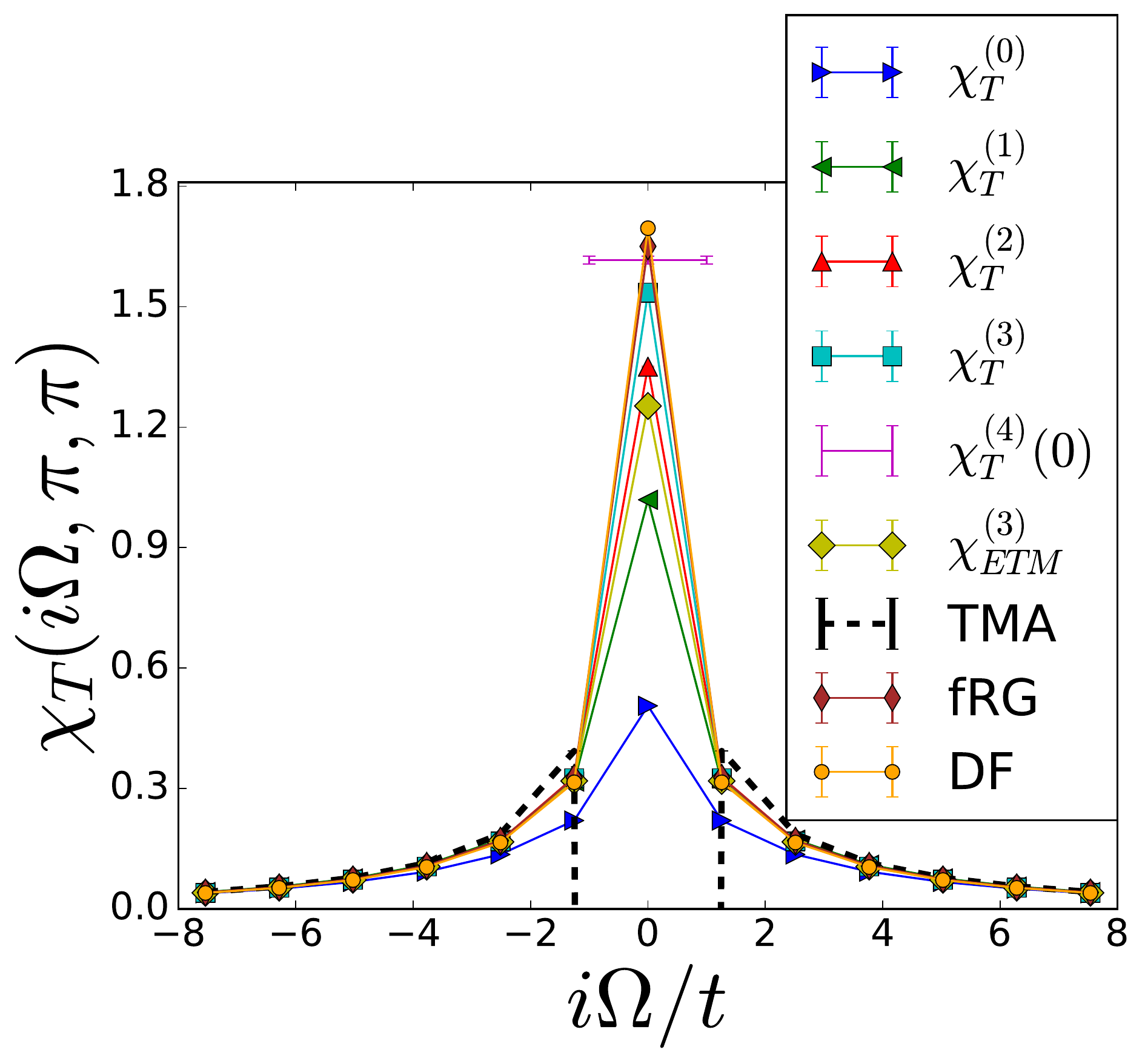}
\caption{ Transverse spin susceptibility $\chi_T^{(m_c)}$ for $m_c = 0,1,2$ and $3$ and $\chi_{ETM}^{(3)}$  vs. Matsubara frequency $i\Omega$ at $U/t=2$, $\beta t=5$ and $\mu/t=0$ for $\textbf q = (\pi,\pi)$.  We also present $\chi_T^{(4)}(i\Omega=0)$, TMA, DF \cite{behnam:2020}, and fRG results from Ref.~\cite{Hille:2020}. \label{fig:chi_vs_mats_freq_beta5}}
\end{figure}
We first consider the order-by-order evaluation of $\chi_T$ on the Matsubara axis at $\textbf{q}=(\pi,\pi)$. Results for truncation order $m_c=0$ to $4$ are shown in Fig.~\ref{fig:chi_vs_mats_freq_beta5}. 
We consider a weak coupling parameter regime which has been of interest for algorithm development due to the long correlation length of the model, a fact that necessitates very careful finite size scaling for many numerical methods \cite{benchmarks,Schaefer:2020}.
For comparison we include high-quality results from functional-Renormalization Group (fRG) from Ref.~\cite{Hille:2020} and our results from the dual-fermion (DF) technique \cite{behnam:2020}, as well as the evaluation of the TMA.  
The parameter regime of $U/t=2$ at $\beta t=5$ has been chosen precisely because it is the cusp where $U\chi_T^{(0)} \approx 1$ and the TMA breaks down resulting in a diverging negative value at $i\Omega=0$ while the result at all other frequencies is overestimated by the TMA.  In contrast, the order by order expansion is exactly equivalent to the reference data at $i\Omega \neq 0$ and shows a systematic tendency at $i\Omega=0$ towards the reference fRG and DF data sets. By truncation order $m_c=4$ the discrepancy of $\chi_T^{(4)}$  compared to the fRG and DF results  is $\approx 2$\% and 5\% respectively. Also shown are results for $\chi_{ETM}^{(3)}$ which is in precise agreement with the fRG and DF results for $i\Omega\neq 0$.  At $i\Omega=0$ the third-order ETM result does not suffer the divergence of the TMA although it underestimates the value even in comparison to the direct second order expansion.  It seems that the infinite resummation of non-ladder-like diagrams included in the ETM approach does not contain new information, however it provides a path to an approximate solution while evaluating fewer diagrams.
\begin{figure}
\centering
\includegraphics[width=1.0\linewidth]{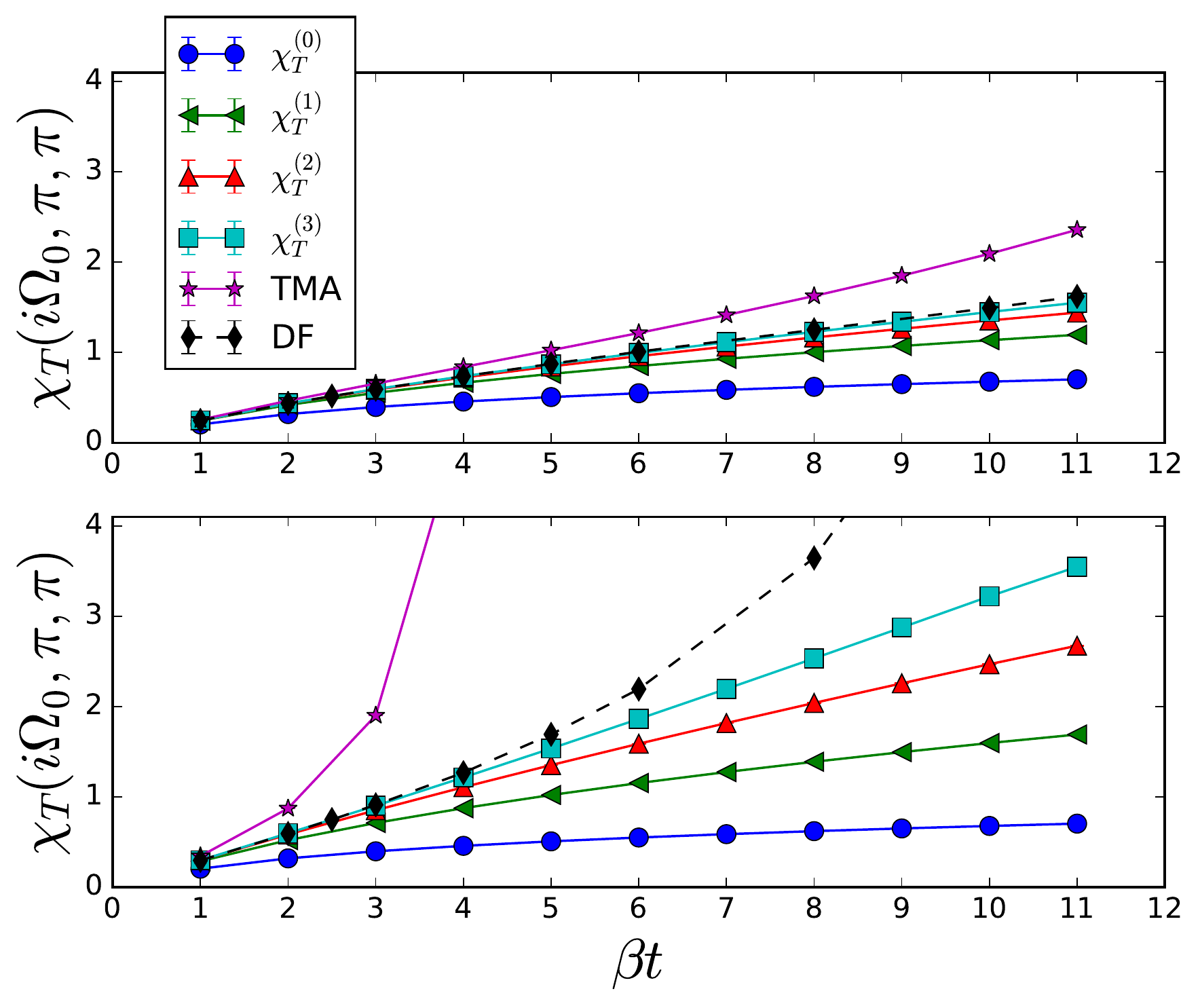}
\caption{ Transverse spin susceptibility vs. inverse temperature at different truncation orders $m_c = 0,1,2$ and $3$. The TMA and DF results are also shown for comparison. Data are for \emph{Top}: $U/t=1$, \emph{Bottom}: $U/t=2$ with $\mu/t=0$ at $\textbf q = (\pi,\pi)$ and $i\Omega = i\Omega_0=0$. \label{fig:chi_vs_beta_omg0}}
\end{figure}

Having now verified the precise convergence at a nominal temperature of $\beta t=5$ we display the order-by-order temperature dependence of the direct expansion, $\chi_T^{(m_c)}$ at $\textbf q=(\pi,\pi)$ for the zeroth bosonic frequency at weak coupling.  Results are shown in Fig.~\ref{fig:chi_vs_beta_omg0} for $U/t=1$ and $2$. 
We include  results of the dual-fermion (DF) method \cite{behnam:2020, opendf, LeBlanc:2019}, which for this parameter range is essentially exact \cite{Gukelberger:DF}, as well as comparison to the TMA.
One immediately notes the deviation of the TMA result from the DF benchmark even at $U/t=1$ for temperatures above $\beta t=2$, which translates into a severe divergence for  $U/t=2$  above $\beta t=1$.  The TMA therefore has an extremely limited range of applicability within condensed matter systems even for very weak interactions and high temperatures.  In contrast, the order-by-order expansion remains stable, showing a systematic improvement,  and we see that higher orders become more important at lower temperatures and larger $U/t$ values.  
The data point $\textbf q=(\pi,\pi)$ and $i\Omega=i\Omega_0$
is the point where the convergence of the transverse susceptibility series is slowest.
However, for non-zero Matsubara frequencies the convergence of the
series is extremely fast, often by second or third order (see the
Supplemental Materials for a non-zero frequency comparison).

\begin{figure}
\centering
\includegraphics[width=1.00\linewidth]{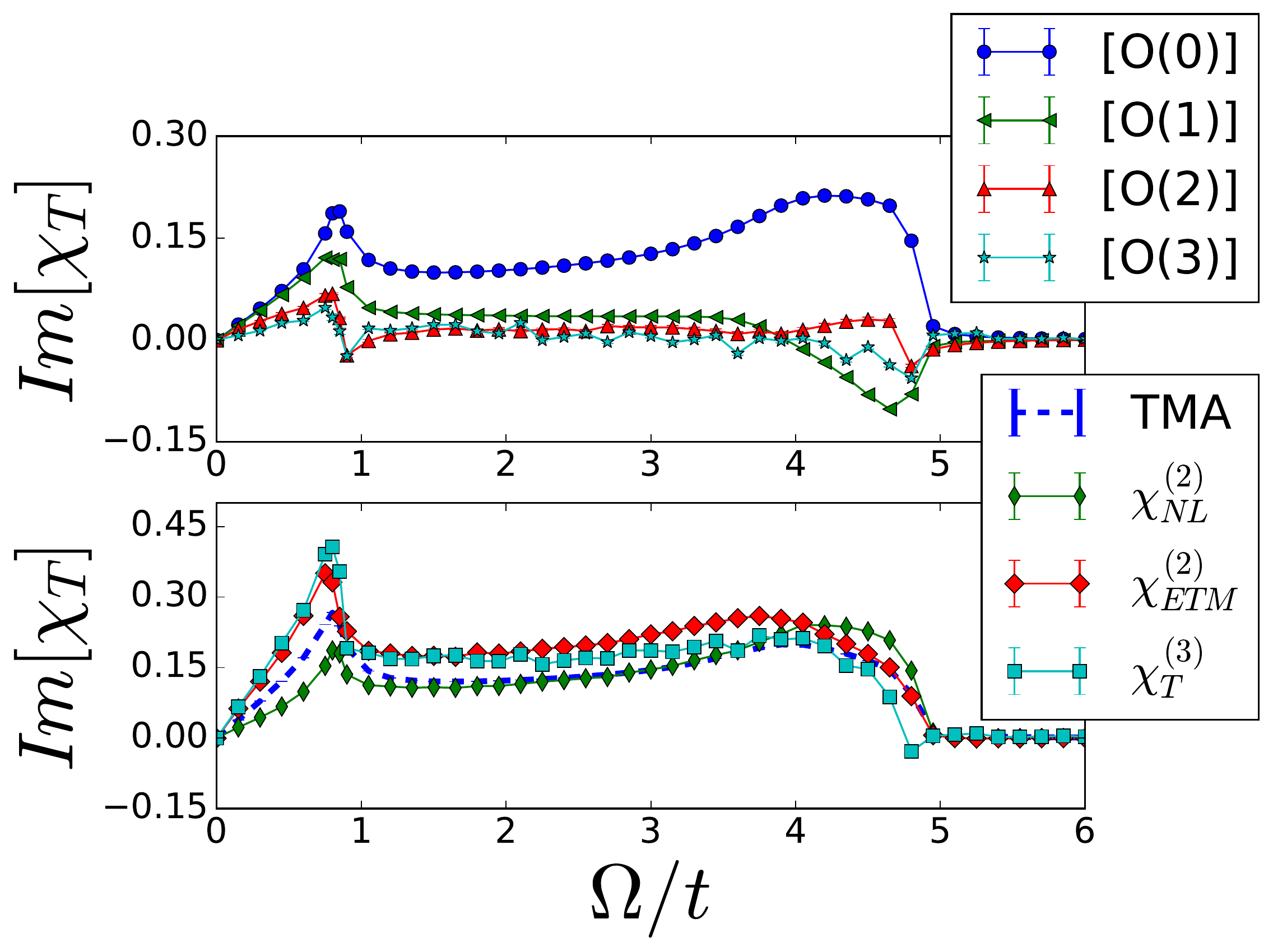}
\caption{ \emph{Top}: Imaginary part of the $m$th order transverse spin susceptibility diagrams $[O(m)]$ vs. real frequency $\Omega$. \emph{Bottom}: The third order transverse spin susceptibility $\chi_T^{(3)}$; TMA, $\chi_{ETM}^{(2)}$  as well as $\chi_{NL}^{(2)}$, are also shown. Data are for $\beta t=5$, $U/t=2$ with $\mu/t=0$ at $\textbf q = (\pi/3,\pi/2)$. We set $\Gamma/t=0.02$ in the symbolic analytic continuation $i\Omega \to \Omega + i\Gamma$. \label{fig:chi_real_freq_order_by_order}}
\end{figure}

We now turn to one of our main results, the order-by-order contribution
of diagrams $[O(m)]$ with $m = 0,1,2,$ and $3$ to the imaginary part of the
transverse spin susceptibility as a function of real frequency (see
Supplemental Materials for ${\rm Re}[\chi_T]$ results).
The top frame of Fig.~\ref{fig:chi_real_freq_order_by_order} shows the
contribution to $\chi_T^{(3)}$ at each separate order.
As the order increases we find that higher order terms
contribute significant corrections
only for a range of frequencies near $\Omega=0$ which adjust the slope
of the $\omega\to 0$ limit of $\chi_T$.  Otherwise the contributions are
largely unstructured until one reaches the band edge near $\omega/t=4$.
The reduced contribution at higher frequency is expected and is similar
to that seen on the Matsubara axis.

Recall that the ETM approximation originates from the evaluation of a
subset of diagrams with non-ladder-like structure $\chi_{NL}^{(m)}$, the
result of which is inverted using Eq.~(\ref{E:chiT_from_chi0}).
We produce $\chi_{ETM}^{(2)}$ for real-frequencies using $\chi_{NL}^{(2)}$ and compare this
in the lower frame of Fig.~\ref{fig:chi_real_freq_order_by_order} to the TMA and the direct
expansion up to third order $\chi_{T}^{(3)}$.
We see that, for this value of $U/t$, the TMA underestimates both the peak amplitude and low frequency slope (and performs worse for larger values of $U/t$ - see Supplemental Materials).
In contrast $\chi_{ETM}^{(2)}$ is nearly identical to the third order direct
expansion - it captures the same slope at low frequency, peak location,
and high frequency amplitude. 
This comparison is rather impressive given
that the third order direct expansion includes 23 diagrams while the ETM
at second order includes only 4. The computational savings created by
reducing the diagrammatic space comes at the cost that the uncertainty
in the ETM approximation is very sensitive to the inversion of
Eq.~(\ref{E:chiT_from_chi0}) and therefore requires high precision
results in order to maintain accuracy.
\begin{figure}
\centering
\includegraphics[width=1.00\linewidth]{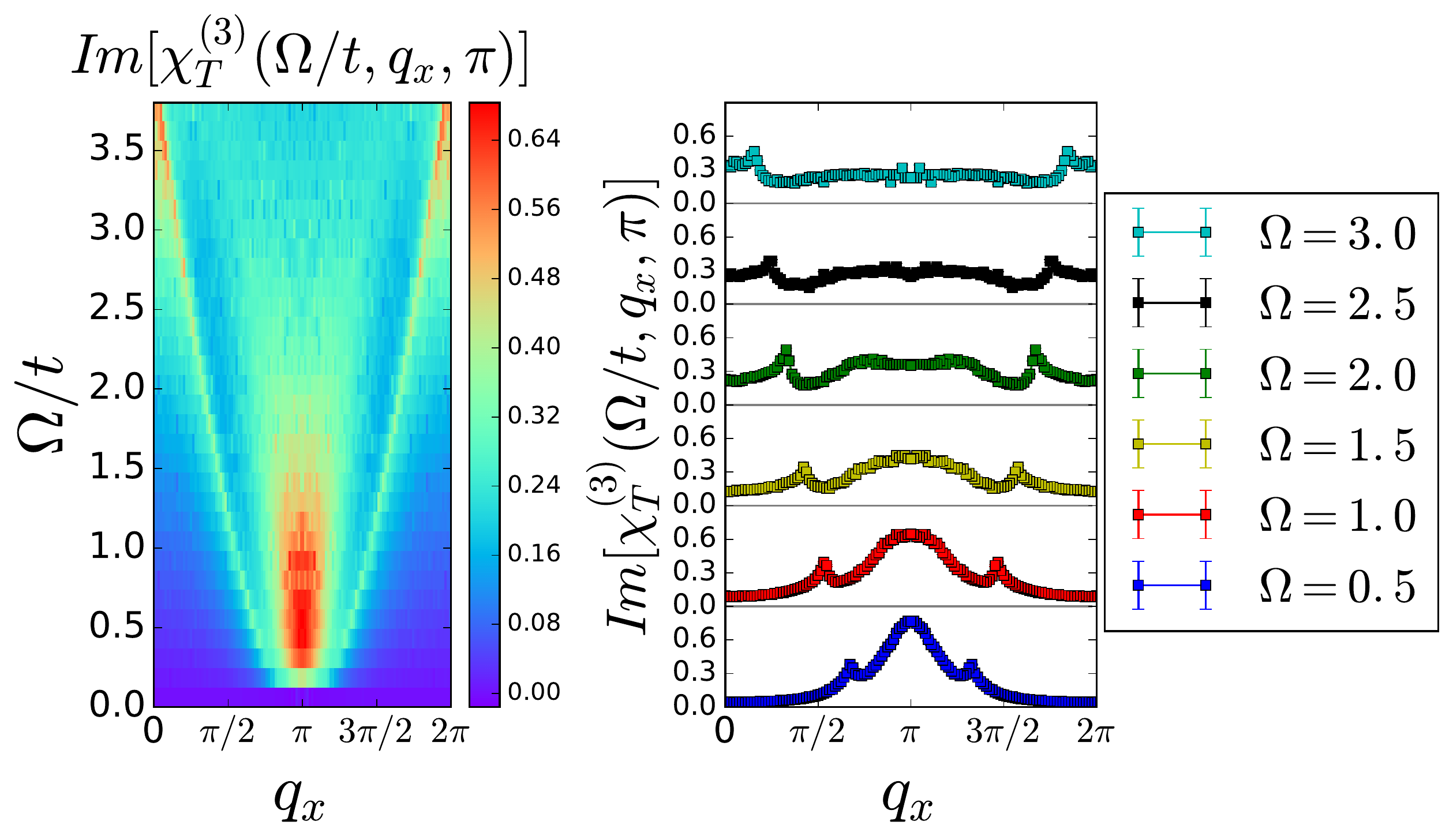}
\caption{ \emph{Left}: third order transverse susceptibility $\chi_T^{(3)}$ as a function of real frequency $\Omega$ along the momentum cut $(0,\pi) \to (2\pi,\pi)$. \emph{Right}: Frequency cuts along $\Omega=0.5$ to $3$. Data are for $\beta t=5$, $U/t=2$, and $\mu/t=0$. We set $\Gamma/t=0.02$ in the symbolic analytic continuation $i\Omega \to \Omega + i\Gamma$. \label{fig:colored_map_U2_two_rows} }
\end{figure}

Finally, we show on the left in Fig.~\ref{fig:colored_map_U2_two_rows} the dependence of the third order spin susceptibility on both $\textbf q$ and real frequency, $\Omega$, along $\textbf q=(0,\pi)$ to $\textbf q=(2\pi,\pi)$ in the first Brillouin zone.  On the right we plot the corresponding susceptibility along several fixed-frequency cuts ranging from  $\Omega/t=0.5$ to $3$.  One notes two important features: a set of two dispersive peaks and a broad peak near $\textbf q=(\pi,\pi)$ that widens and flattens as frequency is increased.   
This behavior is reminiscent of inelastic neutron scattering results on undoped LSCO \cite{hayden:1990,headings:2010} where low energy cuts exhibit a single peak near $\textbf q=(\pi,\pi)$ that splits at higher energies into a set of two dispersive peaks.  Those materials have been understood, however, with linear-spin-wave models that estimate values of $U/t\approx 8$, well beyond the convergence of our series at low orders \cite{coldea:2001}.  Precisely how spin excitations evolve from weak to strong coupling in Hubbard models has yet to be understood.  Our method might be extended to larger values of $U/t$ via renormalization procedures, but such work has yet to be accomplished.

\indent \emph{Concluding remarks:}
We have computed a direct perturbative diagrammatic expansion of the spin susceptibility {\em
evaluated on the real frequency axis} facilitated by AMI.
We point out that there is no conceptual hurdle in generating and evaluating
higher order diagrams. 
AMI automates this process and provides an {\em analytic} result in
frequency space that can
be expressed on the real frequency axis by simple substitution without
resorting to numerical analytic continuation methods.
The standard methods of numerical analytic continuation (such as MAXENT
or pade-approximants), while known to be ill-posed, have been
central in the theoretical analysis of both single and two-particle
properties of materials.
With the advent of AMI, this no longer need be the case for any problem
where direct perturbative expansions are convergent.

This methodology has the advantage that is it conceptually very simple and appears to be systematically controllable. The analytic expressions generated by AMI to solve the 2D square lattice
with Hubbard interaction remain valid in any dimensionality for any
single-band dispersion. Moreover, the AMI procedure is not limited to
Hubbard interactions and can be applied to any frequency independent
interaction \cite{AMI,GIT}. The procedures outlined in this work can therefore be applied to the diagrammatic expansion of the polarization function relevant to screening problems such as the GW approximation. 
We anticipate that the application of AMI to other
interactions on lattice systems will open many avenues of advancement in
condensed matter physics.

The authors would like to thank  F. \ifmmode \check{S}\else \v{S}\fi{}imkovic, S. Andergassen, C. Hille and T. Schäfer for reference data and useful discussion. JPFL and SHC acknowledge the support of the Natural Sciences and Engineering Research Council of Canada (NSERC) (RGPIN-2017-04253, and RGPIN-2014-057). 
Computational resources were provided by ACENET and Compute Canada. Our Monte Carlo codes make use of the open source ALPSCore framework \cite{Gaenko17,ALPSCore,alpscore_v2} and we have used the open source code Maxent \cite{Levy2016} for numerical analytic continuation.

\bibliographystyle{apsrev4-1}
\bibliography{refs.bib}

\begin{thebibliography}{65}%
\makeatletter
\providecommand \@ifxundefined [1]{%
 \@ifx{#1\undefined}
}%
\providecommand \@ifnum [1]{%
 \ifnum #1\expandafter \@firstoftwo
 \else \expandafter \@secondoftwo
 \fi
}%
\providecommand \@ifx [1]{%
 \ifx #1\expandafter \@firstoftwo
 \else \expandafter \@secondoftwo
 \fi
}%
\providecommand \natexlab [1]{#1}%
\providecommand \enquote  [1]{``#1''}%
\providecommand \bibnamefont  [1]{#1}%
\providecommand \bibfnamefont [1]{#1}%
\providecommand \citenamefont [1]{#1}%
\providecommand \href@noop [0]{\@secondoftwo}%
\providecommand \href [0]{\begingroup \@sanitize@url \@href}%
\providecommand \@href[1]{\@@startlink{#1}\@@href}%
\providecommand \@@href[1]{\endgroup#1\@@endlink}%
\providecommand \@sanitize@url [0]{\catcode `\\12\catcode `\$12\catcode
  `\&12\catcode `\#12\catcode `\^12\catcode `\_12\catcode `\%12\relax}%
\providecommand \@@startlink[1]{}%
\providecommand \@@endlink[0]{}%
\providecommand \url  [0]{\begingroup\@sanitize@url \@url }%
\providecommand \@url [1]{\endgroup\@href {#1}{\urlprefix }}%
\providecommand \urlprefix  [0]{URL }%
\providecommand \Eprint [0]{\href }%
\providecommand \doibase [0]{http://dx.doi.org/}%
\providecommand \selectlanguage [0]{\@gobble}%
\providecommand \bibinfo  [0]{\@secondoftwo}%
\providecommand \bibfield  [0]{\@secondoftwo}%
\providecommand \translation [1]{[#1]}%
\providecommand \BibitemOpen [0]{}%
\providecommand \bibitemStop [0]{}%
\providecommand \bibitemNoStop [0]{.\EOS\space}%
\providecommand \EOS [0]{\spacefactor3000\relax}%
\providecommand \BibitemShut  [1]{\csname bibitem#1\endcsname}%
\let\auto@bib@innerbib\@empty
\bibitem [{\citenamefont {Hubbard}(1963)}]{Hubbard1963}%
  \BibitemOpen
  \bibfield  {author} {\bibinfo {author} {\bibfnamefont {J.}~\bibnamefont
  {Hubbard}},\ }\href
  {http://rspa.royalsocietypublishing.org/content/276/1365/238.abstract}
  {\bibfield  {journal} {\bibinfo  {journal} {Proc. R. Soc. London, Ser. A}\
  }\textbf {\bibinfo {volume} {276}},\ \bibinfo {pages} {238} (\bibinfo {year}
  {1963})}\BibitemShut {NoStop}%
\bibitem [{\citenamefont {Mancini}\ \emph {et~al.}(1995)\citenamefont
  {Mancini}, \citenamefont {Marra},\ and\ \citenamefont
  {Matsumoto}}]{MANCINI1995361}%
  \BibitemOpen
  \bibfield  {author} {\bibinfo {author} {\bibfnamefont {F.}~\bibnamefont
  {Mancini}}, \bibinfo {author} {\bibfnamefont {S.}~\bibnamefont {Marra}}, \
  and\ \bibinfo {author} {\bibfnamefont {H.}~\bibnamefont {Matsumoto}},\ }\href
  {\doibase https://doi.org/10.1016/0921-4534(95)00482-3} {\bibfield  {journal}
  {\bibinfo  {journal} {Physica C: Superconductivity}\ }\textbf {\bibinfo
  {volume} {252}},\ \bibinfo {pages} {361 } (\bibinfo {year}
  {1995})}\BibitemShut {NoStop}%
\bibitem [{\citenamefont {LeBlanc}\ \emph {et~al.}(2015)\citenamefont
  {LeBlanc}, \citenamefont {Antipov}, \citenamefont {Becca}, \citenamefont
  {Bulik}, \citenamefont {Chan}, \citenamefont {Chung}, \citenamefont {Deng},
  \citenamefont {Ferrero}, \citenamefont {Henderson}, \citenamefont
  {Jim\'enez-Hoyos}, \citenamefont {Kozik}, \citenamefont {Liu}, \citenamefont
  {Millis}, \citenamefont {Prokof'ev}, \citenamefont {Qin}, \citenamefont
  {Scuseria}, \citenamefont {Shi}, \citenamefont {Svistunov}, \citenamefont
  {Tocchio}, \citenamefont {Tupitsyn}, \citenamefont {White}, \citenamefont
  {Zhang}, \citenamefont {Zheng}, \citenamefont {Zhu},\ and\ \citenamefont
  {Gull}}]{benchmarks}%
  \BibitemOpen
  \bibfield  {author} {\bibinfo {author} {\bibfnamefont {J.~P.~F.}\
  \bibnamefont {LeBlanc}}, \bibinfo {author} {\bibfnamefont {A.~E.}\
  \bibnamefont {Antipov}}, \bibinfo {author} {\bibfnamefont {F.}~\bibnamefont
  {Becca}}, \bibinfo {author} {\bibfnamefont {I.~W.}\ \bibnamefont {Bulik}},
  \bibinfo {author} {\bibfnamefont {G.~K.-L.}\ \bibnamefont {Chan}}, \bibinfo
  {author} {\bibfnamefont {C.-M.}\ \bibnamefont {Chung}}, \bibinfo {author}
  {\bibfnamefont {Y.}~\bibnamefont {Deng}}, \bibinfo {author} {\bibfnamefont
  {M.}~\bibnamefont {Ferrero}}, \bibinfo {author} {\bibfnamefont {T.~M.}\
  \bibnamefont {Henderson}}, \bibinfo {author} {\bibfnamefont {C.~A.}\
  \bibnamefont {Jim\'enez-Hoyos}}, \bibinfo {author} {\bibfnamefont
  {E.}~\bibnamefont {Kozik}}, \bibinfo {author} {\bibfnamefont {X.-W.}\
  \bibnamefont {Liu}}, \bibinfo {author} {\bibfnamefont {A.~J.}\ \bibnamefont
  {Millis}}, \bibinfo {author} {\bibfnamefont {N.~V.}\ \bibnamefont
  {Prokof'ev}}, \bibinfo {author} {\bibfnamefont {M.}~\bibnamefont {Qin}},
  \bibinfo {author} {\bibfnamefont {G.~E.}\ \bibnamefont {Scuseria}}, \bibinfo
  {author} {\bibfnamefont {H.}~\bibnamefont {Shi}}, \bibinfo {author}
  {\bibfnamefont {B.~V.}\ \bibnamefont {Svistunov}}, \bibinfo {author}
  {\bibfnamefont {L.~F.}\ \bibnamefont {Tocchio}}, \bibinfo {author}
  {\bibfnamefont {I.~S.}\ \bibnamefont {Tupitsyn}}, \bibinfo {author}
  {\bibfnamefont {S.~R.}\ \bibnamefont {White}}, \bibinfo {author}
  {\bibfnamefont {S.}~\bibnamefont {Zhang}}, \bibinfo {author} {\bibfnamefont
  {B.-X.}\ \bibnamefont {Zheng}}, \bibinfo {author} {\bibfnamefont
  {Z.}~\bibnamefont {Zhu}}, \ and\ \bibinfo {author} {\bibfnamefont
  {E.}~\bibnamefont {Gull}} (\bibinfo {collaboration} {Simons Collaboration on
  the Many-Electron Problem}),\ }\href {\doibase 10.1103/PhysRevX.5.041041}
  {\bibfield  {journal} {\bibinfo  {journal} {Phys. Rev. X}\ }\textbf {\bibinfo
  {volume} {5}},\ \bibinfo {pages} {041041} (\bibinfo {year}
  {2015})}\BibitemShut {NoStop}%
\bibitem [{\citenamefont {Schäfer}\ \emph {et~al.}(2020)\citenamefont
  {Schäfer}, \citenamefont {Wentzell}, \citenamefont {Šimkovic},
  \citenamefont {He}, \citenamefont {Hille}, \citenamefont {Klett},
  \citenamefont {Eckhardt}, \citenamefont {h}, \citenamefont {i}, \citenamefont
  {Régent}, \citenamefont {Kirsch}, \citenamefont {Wang}, \citenamefont {k},
  \citenamefont {Kozik}, \citenamefont {Stepanov}, \citenamefont {Kauch},
  \citenamefont {Andergassen}, \citenamefont {Hansmann}, \citenamefont {Vilk},
  \citenamefont {LeBlanc}, \citenamefont {Zhang}, \citenamefont {Tremblay},
  \citenamefont {Ferrero}, \citenamefont {Parcollet}, ,\ and\ \citenamefont
  {Georges}}]{Schaefer:2020}%
  \BibitemOpen
  \bibfield  {author} {\bibinfo {author} {\bibfnamefont {T.}~\bibnamefont
  {Schäfer}}, \bibinfo {author} {\bibfnamefont {N.}~\bibnamefont {Wentzell}},
  \bibinfo {author} {\bibfnamefont {F.}~\bibnamefont {Šimkovic}}, \bibinfo
  {author} {\bibfnamefont {Y.-Y.}\ \bibnamefont {He}}, \bibinfo {author}
  {\bibfnamefont {C.}~\bibnamefont {Hille}}, \bibinfo {author} {\bibfnamefont
  {M.}~\bibnamefont {Klett}}, \bibinfo {author} {\bibfnamefont {C.~J.}\
  \bibnamefont {Eckhardt}}, \bibinfo {author} {\bibfnamefont {B.~A.}\
  \bibnamefont {h}}, \bibinfo {author} {\bibfnamefont {V.~H.}\ \bibnamefont
  {i}}, \bibinfo {author} {\bibfnamefont {F.-M.~L.}\ \bibnamefont {Régent}},
  \bibinfo {author} {\bibfnamefont {A.}~\bibnamefont {Kirsch}}, \bibinfo
  {author} {\bibfnamefont {Y.}~\bibnamefont {Wang}}, \bibinfo {author}
  {\bibfnamefont {A.~K.}\ \bibnamefont {k}}, \bibinfo {author} {\bibfnamefont
  {E.}~\bibnamefont {Kozik}}, \bibinfo {author} {\bibfnamefont {E.~A.}\
  \bibnamefont {Stepanov}}, \bibinfo {author} {\bibfnamefont {A.}~\bibnamefont
  {Kauch}}, \bibinfo {author} {\bibfnamefont {S.}~\bibnamefont {Andergassen}},
  \bibinfo {author} {\bibfnamefont {P.}~\bibnamefont {Hansmann}}, \bibinfo
  {author} {\bibfnamefont {Y.~M.}\ \bibnamefont {Vilk}}, \bibinfo {author}
  {\bibfnamefont {J.}~\bibnamefont {LeBlanc}}, \bibinfo {author} {\bibfnamefont
  {S.}~\bibnamefont {Zhang}}, \bibinfo {author} {\bibfnamefont {A.-M.}\
  \bibnamefont {Tremblay}}, \bibinfo {author} {\bibfnamefont {M.}~\bibnamefont
  {Ferrero}}, \bibinfo {author} {\bibfnamefont {O.}~\bibnamefont {Parcollet}},
  , \ and\ \bibinfo {author} {\bibfnamefont {A.}~\bibnamefont {Georges}},\
  }\href@noop {} {\enquote {\bibinfo {title} {Putting modern many-body
  computations to the test: a multi-messenger, multi-method study of the
  half-filled two-dimensional hubbard model at weak coupling},}\ } (\bibinfo
  {year} {2020}),\ \bibinfo {note} {in preparation}\BibitemShut {NoStop}%
\bibitem [{\citenamefont {Kotliar}\ \emph {et~al.}(2001)\citenamefont
  {Kotliar}, \citenamefont {Savrasov}, \citenamefont {P\'alsson},\ and\
  \citenamefont {Biroli}}]{Kotliar01}%
  \BibitemOpen
  \bibfield  {author} {\bibinfo {author} {\bibfnamefont {G.}~\bibnamefont
  {Kotliar}}, \bibinfo {author} {\bibfnamefont {S.~Y.}\ \bibnamefont
  {Savrasov}}, \bibinfo {author} {\bibfnamefont {G.}~\bibnamefont {P\'alsson}},
  \ and\ \bibinfo {author} {\bibfnamefont {G.}~\bibnamefont {Biroli}},\ }\href
  {\doibase 10.1103/PhysRevLett.87.186401} {\bibfield  {journal} {\bibinfo
  {journal} {Phys. Rev. Lett.}\ }\textbf {\bibinfo {volume} {87}},\ \bibinfo
  {pages} {186401} (\bibinfo {year} {2001})}\BibitemShut {NoStop}%
\bibitem [{\citenamefont {Kotliar}\ \emph {et~al.}(2006)\citenamefont
  {Kotliar}, \citenamefont {Savrasov}, \citenamefont {Haule}, \citenamefont
  {Oudovenko}, \citenamefont {Parcollet},\ and\ \citenamefont
  {Marianetti}}]{Kotliar:2006}%
  \BibitemOpen
  \bibfield  {author} {\bibinfo {author} {\bibfnamefont {G.}~\bibnamefont
  {Kotliar}}, \bibinfo {author} {\bibfnamefont {S.~Y.}\ \bibnamefont
  {Savrasov}}, \bibinfo {author} {\bibfnamefont {K.}~\bibnamefont {Haule}},
  \bibinfo {author} {\bibfnamefont {V.~S.}\ \bibnamefont {Oudovenko}}, \bibinfo
  {author} {\bibfnamefont {O.}~\bibnamefont {Parcollet}}, \ and\ \bibinfo
  {author} {\bibfnamefont {C.~A.}\ \bibnamefont {Marianetti}},\ }\href
  {\doibase 10.1103/RevModPhys.78.865} {\bibfield  {journal} {\bibinfo
  {journal} {Rev. Mod. Phys.}\ }\textbf {\bibinfo {volume} {78}},\ \bibinfo
  {pages} {865} (\bibinfo {year} {2006})}\BibitemShut {NoStop}%
\bibitem [{\citenamefont {Hettler}\ \emph {et~al.}(1998)\citenamefont
  {Hettler}, \citenamefont {Tahvildar-Zadeh}, \citenamefont {Jarrell},
  \citenamefont {Pruschke},\ and\ \citenamefont
  {Krishnamurthy}}]{hettler:1998}%
  \BibitemOpen
  \bibfield  {author} {\bibinfo {author} {\bibfnamefont {M.~H.}\ \bibnamefont
  {Hettler}}, \bibinfo {author} {\bibfnamefont {A.~N.}\ \bibnamefont
  {Tahvildar-Zadeh}}, \bibinfo {author} {\bibfnamefont {M.}~\bibnamefont
  {Jarrell}}, \bibinfo {author} {\bibfnamefont {T.}~\bibnamefont {Pruschke}}, \
  and\ \bibinfo {author} {\bibfnamefont {H.~R.}\ \bibnamefont
  {Krishnamurthy}},\ }\href {\doibase 10.1103/PhysRevB.58.R7475} {\bibfield
  {journal} {\bibinfo  {journal} {Phys. Rev. B}\ }\textbf {\bibinfo {volume}
  {58}},\ \bibinfo {pages} {R7475} (\bibinfo {year} {1998})}\BibitemShut
  {NoStop}%
\bibitem [{\citenamefont {Hettler}\ \emph {et~al.}(2000)\citenamefont
  {Hettler}, \citenamefont {Mukherjee}, \citenamefont {Jarrell},\ and\
  \citenamefont {Krishnamurthy}}]{hettler:2000}%
  \BibitemOpen
  \bibfield  {author} {\bibinfo {author} {\bibfnamefont {M.~H.}\ \bibnamefont
  {Hettler}}, \bibinfo {author} {\bibfnamefont {M.}~\bibnamefont {Mukherjee}},
  \bibinfo {author} {\bibfnamefont {M.}~\bibnamefont {Jarrell}}, \ and\
  \bibinfo {author} {\bibfnamefont {H.~R.}\ \bibnamefont {Krishnamurthy}},\
  }\href {\doibase 10.1103/PhysRevB.61.12739} {\bibfield  {journal} {\bibinfo
  {journal} {Phys. Rev. B}\ }\textbf {\bibinfo {volume} {61}},\ \bibinfo
  {pages} {12739} (\bibinfo {year} {2000})}\BibitemShut {NoStop}%
\bibitem [{\citenamefont {Prokof'ev}\ and\ \citenamefont
  {Svistunov}(1998)}]{Prokof'ev:1998}%
  \BibitemOpen
  \bibfield  {author} {\bibinfo {author} {\bibfnamefont {N.~V.}\ \bibnamefont
  {Prokof'ev}}\ and\ \bibinfo {author} {\bibfnamefont {B.~V.}\ \bibnamefont
  {Svistunov}},\ }\href {\doibase 10.1103/PhysRevLett.81.2514} {\bibfield
  {journal} {\bibinfo  {journal} {Phys. Rev. Lett.}\ }\textbf {\bibinfo
  {volume} {81}},\ \bibinfo {pages} {2514} (\bibinfo {year}
  {1998})}\BibitemShut {NoStop}%
\bibitem [{\citenamefont {Houcke}\ \emph {et~al.}(2010)\citenamefont {Houcke},
  \citenamefont {Kozik}, \citenamefont {Prokof’ev},\ and\ \citenamefont
  {Svistunov}}]{vanhoucke}%
  \BibitemOpen
  \bibfield  {author} {\bibinfo {author} {\bibfnamefont {K.~V.}\ \bibnamefont
  {Houcke}}, \bibinfo {author} {\bibfnamefont {E.}~\bibnamefont {Kozik}},
  \bibinfo {author} {\bibfnamefont {N.}~\bibnamefont {Prokof’ev}}, \ and\
  \bibinfo {author} {\bibfnamefont {B.}~\bibnamefont {Svistunov}},\ }\href
  {\doibase https://doi.org/10.1016/j.phpro.2010.09.034} {\bibfield  {journal}
  {\bibinfo  {journal} {Physics Procedia}\ }\textbf {\bibinfo {volume} {6}},\
  \bibinfo {pages} {95 } (\bibinfo {year} {2010})}\BibitemShut {NoStop}%
\bibitem [{\citenamefont {{Van Houcke}}\ \emph {et~al.}(2012)\citenamefont
  {{Van Houcke}}, \citenamefont {Werner}, \citenamefont {Kozik}, \citenamefont
  {{Prokof'ev}}, \citenamefont {Svistunov}, \citenamefont {Ku}, \citenamefont
  {Sommer}, \citenamefont {Cheuk}, \citenamefont {Schirotzek},\ and\
  \citenamefont {Zwierlein}}]{vanhoucke:natphys}%
  \BibitemOpen
  \bibfield  {author} {\bibinfo {author} {\bibfnamefont {K.}~\bibnamefont {{Van
  Houcke}}}, \bibinfo {author} {\bibfnamefont {F.}~\bibnamefont {Werner}},
  \bibinfo {author} {\bibfnamefont {E.}~\bibnamefont {Kozik}}, \bibinfo
  {author} {\bibfnamefont {N.}~\bibnamefont {{Prokof'ev}}}, \bibinfo {author}
  {\bibfnamefont {B.}~\bibnamefont {Svistunov}}, \bibinfo {author}
  {\bibfnamefont {M.~J.~H.}\ \bibnamefont {Ku}}, \bibinfo {author}
  {\bibfnamefont {A.~T.}\ \bibnamefont {Sommer}}, \bibinfo {author}
  {\bibfnamefont {L.~W.}\ \bibnamefont {Cheuk}}, \bibinfo {author}
  {\bibfnamefont {A.}~\bibnamefont {Schirotzek}}, \ and\ \bibinfo {author}
  {\bibfnamefont {M.~W.}\ \bibnamefont {Zwierlein}},\ }\href@noop {} {\bibfield
   {journal} {\bibinfo  {journal} {Nature Physics}\ }\textbf {\bibinfo {volume}
  {8}},\ \bibinfo {pages} {366} (\bibinfo {year} {2012})}\BibitemShut {NoStop}%
\bibitem [{\citenamefont {Kozik}\ \emph {et~al.}(2010)\citenamefont {Kozik},
  \citenamefont {Houcke}, \citenamefont {Gull}, \citenamefont {Pollet},
  \citenamefont {Prokof'ev}, \citenamefont {Svistunov},\ and\ \citenamefont
  {Troyer}}]{kozik:2010}%
  \BibitemOpen
  \bibfield  {author} {\bibinfo {author} {\bibfnamefont {E.}~\bibnamefont
  {Kozik}}, \bibinfo {author} {\bibfnamefont {K.~V.}\ \bibnamefont {Houcke}},
  \bibinfo {author} {\bibfnamefont {E.}~\bibnamefont {Gull}}, \bibinfo {author}
  {\bibfnamefont {L.}~\bibnamefont {Pollet}}, \bibinfo {author} {\bibfnamefont
  {N.}~\bibnamefont {Prokof'ev}}, \bibinfo {author} {\bibfnamefont
  {B.}~\bibnamefont {Svistunov}}, \ and\ \bibinfo {author} {\bibfnamefont
  {M.}~\bibnamefont {Troyer}},\ }\href
  {http://stacks.iop.org/0295-5075/90/i=1/a=10004} {\bibfield  {journal}
  {\bibinfo  {journal} {EPL (Europhysics Letters)}\ }\textbf {\bibinfo {volume}
  {90}},\ \bibinfo {pages} {10004} (\bibinfo {year} {2010})}\BibitemShut
  {NoStop}%
\bibitem [{\citenamefont {Rossi}(2017)}]{rossi2017determinant}%
  \BibitemOpen
  \bibfield  {author} {\bibinfo {author} {\bibfnamefont {R.}~\bibnamefont
  {Rossi}},\ }\href {\doibase 10.1103/PhysRevLett.119.045701} {\bibfield
  {journal} {\bibinfo  {journal} {Phys. Rev. Lett.}\ }\textbf {\bibinfo
  {volume} {119}},\ \bibinfo {pages} {045701} (\bibinfo {year}
  {2017})}\BibitemShut {NoStop}%
\bibitem [{\citenamefont {Rossi}\ \emph {et~al.}(2016)\citenamefont {Rossi},
  \citenamefont {Werner}, \citenamefont {Prokof'ev},\ and\ \citenamefont
  {Svistunov}}]{Rossi:shiftedaction}%
  \BibitemOpen
  \bibfield  {author} {\bibinfo {author} {\bibfnamefont {R.}~\bibnamefont
  {Rossi}}, \bibinfo {author} {\bibfnamefont {F.}~\bibnamefont {Werner}},
  \bibinfo {author} {\bibfnamefont {N.}~\bibnamefont {Prokof'ev}}, \ and\
  \bibinfo {author} {\bibfnamefont {B.}~\bibnamefont {Svistunov}},\ }\href
  {\doibase 10.1103/PhysRevB.93.161102} {\bibfield  {journal} {\bibinfo
  {journal} {Phys. Rev. B}\ }\textbf {\bibinfo {volume} {93}},\ \bibinfo
  {pages} {161102} (\bibinfo {year} {2016})}\BibitemShut {NoStop}%
\bibitem [{\citenamefont {Moutenet}\ \emph {et~al.}(2018)\citenamefont
  {Moutenet}, \citenamefont {Wu},\ and\ \citenamefont
  {Ferrero}}]{ferrero:2018}%
  \BibitemOpen
  \bibfield  {author} {\bibinfo {author} {\bibfnamefont {A.}~\bibnamefont
  {Moutenet}}, \bibinfo {author} {\bibfnamefont {W.}~\bibnamefont {Wu}}, \ and\
  \bibinfo {author} {\bibfnamefont {M.}~\bibnamefont {Ferrero}},\ }\href
  {\doibase 10.1103/PhysRevB.97.085117} {\bibfield  {journal} {\bibinfo
  {journal} {Phys. Rev. B}\ }\textbf {\bibinfo {volume} {97}},\ \bibinfo
  {pages} {085117} (\bibinfo {year} {2018})}\BibitemShut {NoStop}%
\bibitem [{\citenamefont {\ifmmode~\check{S}\else \v{S}\fi{}imkovic}\ and\
  \citenamefont {Kozik}(2019)}]{Evgeny:2019}%
  \BibitemOpen
  \bibfield  {author} {\bibinfo {author} {\bibfnamefont {F.}~\bibnamefont
  {\ifmmode~\check{S}\else \v{S}\fi{}imkovic}}\ and\ \bibinfo {author}
  {\bibfnamefont {E.}~\bibnamefont {Kozik}},\ }\href {\doibase
  10.1103/PhysRevB.100.121102} {\bibfield  {journal} {\bibinfo  {journal}
  {Phys. Rev. B}\ }\textbf {\bibinfo {volume} {100}},\ \bibinfo {pages}
  {121102} (\bibinfo {year} {2019})}\BibitemShut {NoStop}%
\bibitem [{\citenamefont {Takigawa}\ and\ \citenamefont
  {Mitzi}(1994)}]{Takigawa1994}%
  \BibitemOpen
  \bibfield  {author} {\bibinfo {author} {\bibfnamefont {M.}~\bibnamefont
  {Takigawa}}\ and\ \bibinfo {author} {\bibfnamefont {D.~B.}\ \bibnamefont
  {Mitzi}},\ }\href {\doibase 10.1007/BF00754926} {\bibfield  {journal}
  {\bibinfo  {journal} {Journal of Low Temperature Physics}\ }\textbf {\bibinfo
  {volume} {95}},\ \bibinfo {pages} {89} (\bibinfo {year} {1994})}\BibitemShut
  {NoStop}%
\bibitem [{\citenamefont {Coldea}\ \emph {et~al.}(2001)\citenamefont {Coldea},
  \citenamefont {Hayden}, \citenamefont {Aeppli}, \citenamefont {Perring},
  \citenamefont {Frost}, \citenamefont {Mason}, \citenamefont {Cheong},\ and\
  \citenamefont {Fisk}}]{coldea:2001}%
  \BibitemOpen
  \bibfield  {author} {\bibinfo {author} {\bibfnamefont {R.}~\bibnamefont
  {Coldea}}, \bibinfo {author} {\bibfnamefont {S.~M.}\ \bibnamefont {Hayden}},
  \bibinfo {author} {\bibfnamefont {G.}~\bibnamefont {Aeppli}}, \bibinfo
  {author} {\bibfnamefont {T.~G.}\ \bibnamefont {Perring}}, \bibinfo {author}
  {\bibfnamefont {C.~D.}\ \bibnamefont {Frost}}, \bibinfo {author}
  {\bibfnamefont {T.~E.}\ \bibnamefont {Mason}}, \bibinfo {author}
  {\bibfnamefont {S.-W.}\ \bibnamefont {Cheong}}, \ and\ \bibinfo {author}
  {\bibfnamefont {Z.}~\bibnamefont {Fisk}},\ }\href {\doibase
  10.1103/PhysRevLett.86.5377} {\bibfield  {journal} {\bibinfo  {journal}
  {Phys. Rev. Lett.}\ }\textbf {\bibinfo {volume} {86}},\ \bibinfo {pages}
  {5377} (\bibinfo {year} {2001})}\BibitemShut {NoStop}%
\bibitem [{\citenamefont {Fujita}\ \emph {et~al.}(2012)\citenamefont {Fujita},
  \citenamefont {Hiraka}, \citenamefont {Matsuda}, \citenamefont {Matsuura},
  \citenamefont {M.~Tranquada}, \citenamefont {Wakimoto}, \citenamefont {Xu},\
  and\ \citenamefont {Yamada}}]{Fujita:2012}%
  \BibitemOpen
  \bibfield  {author} {\bibinfo {author} {\bibfnamefont {M.}~\bibnamefont
  {Fujita}}, \bibinfo {author} {\bibfnamefont {H.}~\bibnamefont {Hiraka}},
  \bibinfo {author} {\bibfnamefont {M.}~\bibnamefont {Matsuda}}, \bibinfo
  {author} {\bibfnamefont {M.}~\bibnamefont {Matsuura}}, \bibinfo {author}
  {\bibfnamefont {J.}~\bibnamefont {M.~Tranquada}}, \bibinfo {author}
  {\bibfnamefont {S.}~\bibnamefont {Wakimoto}}, \bibinfo {author}
  {\bibfnamefont {G.}~\bibnamefont {Xu}}, \ and\ \bibinfo {author}
  {\bibfnamefont {K.}~\bibnamefont {Yamada}},\ }\href {\doibase
  10.1143/JPSJ.81.011007} {\bibfield  {journal} {\bibinfo  {journal} {Journal
  of the Physical Society of Japan}\ }\textbf {\bibinfo {volume} {81}},\
  \bibinfo {pages} {011007} (\bibinfo {year} {2012})}\BibitemShut {NoStop}%
\bibitem [{\citenamefont {Suzuki}\ \emph {et~al.}(2018)\citenamefont {Suzuki},
  \citenamefont {Minola}, \citenamefont {Lu}, \citenamefont {Peng},
  \citenamefont {Fumagalli}, \citenamefont {Lefran{\c{c}}ois}, \citenamefont
  {Loew}, \citenamefont {Porras}, \citenamefont {Kummer}, \citenamefont
  {Betto}, \citenamefont {Ishida}, \citenamefont {Eisaki}, \citenamefont {Hu},
  \citenamefont {Zhou}, \citenamefont {Haverkort}, \citenamefont {Brookes},
  \citenamefont {Braicovich}, \citenamefont {Ghiringhelli}, \citenamefont
  {Le~Tacon},\ and\ \citenamefont {Keimer}}]{Suzuki2018}%
  \BibitemOpen
  \bibfield  {author} {\bibinfo {author} {\bibfnamefont {H.}~\bibnamefont
  {Suzuki}}, \bibinfo {author} {\bibfnamefont {M.}~\bibnamefont {Minola}},
  \bibinfo {author} {\bibfnamefont {Y.}~\bibnamefont {Lu}}, \bibinfo {author}
  {\bibfnamefont {Y.}~\bibnamefont {Peng}}, \bibinfo {author} {\bibfnamefont
  {R.}~\bibnamefont {Fumagalli}}, \bibinfo {author} {\bibfnamefont
  {E.}~\bibnamefont {Lefran{\c{c}}ois}}, \bibinfo {author} {\bibfnamefont
  {T.}~\bibnamefont {Loew}}, \bibinfo {author} {\bibfnamefont {J.}~\bibnamefont
  {Porras}}, \bibinfo {author} {\bibfnamefont {K.}~\bibnamefont {Kummer}},
  \bibinfo {author} {\bibfnamefont {D.}~\bibnamefont {Betto}}, \bibinfo
  {author} {\bibfnamefont {S.}~\bibnamefont {Ishida}}, \bibinfo {author}
  {\bibfnamefont {H.}~\bibnamefont {Eisaki}}, \bibinfo {author} {\bibfnamefont
  {C.}~\bibnamefont {Hu}}, \bibinfo {author} {\bibfnamefont {X.}~\bibnamefont
  {Zhou}}, \bibinfo {author} {\bibfnamefont {M.~W.}\ \bibnamefont {Haverkort}},
  \bibinfo {author} {\bibfnamefont {N.~B.}\ \bibnamefont {Brookes}}, \bibinfo
  {author} {\bibfnamefont {L.}~\bibnamefont {Braicovich}}, \bibinfo {author}
  {\bibfnamefont {G.}~\bibnamefont {Ghiringhelli}}, \bibinfo {author}
  {\bibfnamefont {M.}~\bibnamefont {Le~Tacon}}, \ and\ \bibinfo {author}
  {\bibfnamefont {B.}~\bibnamefont {Keimer}},\ }\href {\doibase
  10.1038/s41535-018-0139-7} {\bibfield  {journal} {\bibinfo  {journal} {npj
  Quantum Materials}\ }\textbf {\bibinfo {volume} {3}},\ \bibinfo {pages} {65}
  (\bibinfo {year} {2018})}\BibitemShut {NoStop}%
\bibitem [{\citenamefont {Greco}\ \emph {et~al.}(2019)\citenamefont {Greco},
  \citenamefont {Yamase},\ and\ \citenamefont {Bejas}}]{Greco2019}%
  \BibitemOpen
  \bibfield  {author} {\bibinfo {author} {\bibfnamefont {A.}~\bibnamefont
  {Greco}}, \bibinfo {author} {\bibfnamefont {H.}~\bibnamefont {Yamase}}, \
  and\ \bibinfo {author} {\bibfnamefont {M.}~\bibnamefont {Bejas}},\ }\href
  {\doibase 10.1038/s42005-018-0099-z} {\bibfield  {journal} {\bibinfo
  {journal} {Communications Physics}\ }\textbf {\bibinfo {volume} {2}},\
  \bibinfo {pages} {3} (\bibinfo {year} {2019})}\BibitemShut {NoStop}%
\bibitem [{\citenamefont {\ifmmode~\check{S}\else \v{S}\fi{}imkovic}\ \emph
  {et~al.}(2020)\citenamefont {\ifmmode~\check{S}\else \v{S}\fi{}imkovic},
  \citenamefont {LeBlanc}, \citenamefont {Kim}, \citenamefont {Deng},
  \citenamefont {Prokof'ev}, \citenamefont {Svistunov},\ and\ \citenamefont
  {Kozik}}]{fedor:2020}%
  \BibitemOpen
  \bibfield  {author} {\bibinfo {author} {\bibfnamefont {F.}~\bibnamefont
  {\ifmmode~\check{S}\else \v{S}\fi{}imkovic}}, \bibinfo {author}
  {\bibfnamefont {J.~P.~F.}\ \bibnamefont {LeBlanc}}, \bibinfo {author}
  {\bibfnamefont {A.~J.}\ \bibnamefont {Kim}}, \bibinfo {author} {\bibfnamefont
  {Y.}~\bibnamefont {Deng}}, \bibinfo {author} {\bibfnamefont {N.~V.}\
  \bibnamefont {Prokof'ev}}, \bibinfo {author} {\bibfnamefont {B.~V.}\
  \bibnamefont {Svistunov}}, \ and\ \bibinfo {author} {\bibfnamefont
  {E.}~\bibnamefont {Kozik}},\ }\href {\doibase 10.1103/PhysRevLett.124.017003}
  {\bibfield  {journal} {\bibinfo  {journal} {Phys. Rev. Lett.}\ }\textbf
  {\bibinfo {volume} {124}},\ \bibinfo {pages} {017003} (\bibinfo {year}
  {2020})}\BibitemShut {NoStop}%
\bibitem [{\citenamefont {Chen}\ \emph {et~al.}(1994)\citenamefont {Chen},
  \citenamefont {Moreo}, \citenamefont {Ortolani}, \citenamefont {Dagotto},\
  and\ \citenamefont {Lee}}]{Chen:1994}%
  \BibitemOpen
  \bibfield  {author} {\bibinfo {author} {\bibfnamefont {Y.~C.}\ \bibnamefont
  {Chen}}, \bibinfo {author} {\bibfnamefont {A.}~\bibnamefont {Moreo}},
  \bibinfo {author} {\bibfnamefont {F.}~\bibnamefont {Ortolani}}, \bibinfo
  {author} {\bibfnamefont {E.}~\bibnamefont {Dagotto}}, \ and\ \bibinfo
  {author} {\bibfnamefont {T.~K.}\ \bibnamefont {Lee}},\ }\href {\doibase
  10.1103/PhysRevB.50.655} {\bibfield  {journal} {\bibinfo  {journal} {Phys.
  Rev. B}\ }\textbf {\bibinfo {volume} {50}},\ \bibinfo {pages} {655} (\bibinfo
  {year} {1994})}\BibitemShut {NoStop}%
\bibitem [{\citenamefont {Bulut}\ \emph {et~al.}(1995)\citenamefont {Bulut},
  \citenamefont {Scalapino},\ and\ \citenamefont {White}}]{Bulut}%
  \BibitemOpen
  \bibfield  {author} {\bibinfo {author} {\bibfnamefont {N.}~\bibnamefont
  {Bulut}}, \bibinfo {author} {\bibfnamefont {D.}~\bibnamefont {Scalapino}}, \
  and\ \bibinfo {author} {\bibfnamefont {S.}~\bibnamefont {White}},\ }\href
  {\doibase https://doi.org/10.1016/0921-4534(95)00130-1} {\bibfield  {journal}
  {\bibinfo  {journal} {Physica C: Superconductivity}\ }\textbf {\bibinfo
  {volume} {246}},\ \bibinfo {pages} {85 } (\bibinfo {year}
  {1995})}\BibitemShut {NoStop}%
\bibitem [{\citenamefont {Macridin}\ \emph {et~al.}(2006)\citenamefont
  {Macridin}, \citenamefont {Jarrell}, \citenamefont {Maier}, \citenamefont
  {Kent},\ and\ \citenamefont {D'Azevedo}}]{Macridin:2006}%
  \BibitemOpen
  \bibfield  {author} {\bibinfo {author} {\bibfnamefont {A.}~\bibnamefont
  {Macridin}}, \bibinfo {author} {\bibfnamefont {M.}~\bibnamefont {Jarrell}},
  \bibinfo {author} {\bibfnamefont {T.}~\bibnamefont {Maier}}, \bibinfo
  {author} {\bibfnamefont {P.~R.~C.}\ \bibnamefont {Kent}}, \ and\ \bibinfo
  {author} {\bibfnamefont {E.}~\bibnamefont {D'Azevedo}},\ }\href {\doibase
  10.1103/PhysRevLett.97.036401} {\bibfield  {journal} {\bibinfo  {journal}
  {Phys. Rev. Lett.}\ }\textbf {\bibinfo {volume} {97}},\ \bibinfo {pages}
  {036401} (\bibinfo {year} {2006})}\BibitemShut {NoStop}%
\bibitem [{\citenamefont {Gunnarsson}\ \emph {et~al.}(2015)\citenamefont
  {Gunnarsson}, \citenamefont {Sch\"afer}, \citenamefont {LeBlanc},
  \citenamefont {Gull}, \citenamefont {Merino}, \citenamefont {Sangiovanni},
  \citenamefont {Rohringer},\ and\ \citenamefont {Toschi}}]{gunnarsson:2015}%
  \BibitemOpen
  \bibfield  {author} {\bibinfo {author} {\bibfnamefont {O.}~\bibnamefont
  {Gunnarsson}}, \bibinfo {author} {\bibfnamefont {T.}~\bibnamefont
  {Sch\"afer}}, \bibinfo {author} {\bibfnamefont {J.~P.~F.}\ \bibnamefont
  {LeBlanc}}, \bibinfo {author} {\bibfnamefont {E.}~\bibnamefont {Gull}},
  \bibinfo {author} {\bibfnamefont {J.}~\bibnamefont {Merino}}, \bibinfo
  {author} {\bibfnamefont {G.}~\bibnamefont {Sangiovanni}}, \bibinfo {author}
  {\bibfnamefont {G.}~\bibnamefont {Rohringer}}, \ and\ \bibinfo {author}
  {\bibfnamefont {A.}~\bibnamefont {Toschi}},\ }\href {\doibase
  10.1103/PhysRevLett.114.236402} {\bibfield  {journal} {\bibinfo  {journal}
  {Phys. Rev. Lett.}\ }\textbf {\bibinfo {volume} {114}},\ \bibinfo {pages}
  {236402} (\bibinfo {year} {2015})}\BibitemShut {NoStop}%
\bibitem [{\citenamefont {Chen}\ \emph {et~al.}(2015)\citenamefont {Chen},
  \citenamefont {LeBlanc},\ and\ \citenamefont {Gull}}]{Chen:2015}%
  \BibitemOpen
  \bibfield  {author} {\bibinfo {author} {\bibfnamefont {X.}~\bibnamefont
  {Chen}}, \bibinfo {author} {\bibfnamefont {J.~P.~F.}\ \bibnamefont
  {LeBlanc}}, \ and\ \bibinfo {author} {\bibfnamefont {E.}~\bibnamefont
  {Gull}},\ }\href {\doibase 10.1103/PhysRevLett.115.116402} {\bibfield
  {journal} {\bibinfo  {journal} {Phys. Rev. Lett.}\ }\textbf {\bibinfo
  {volume} {115}},\ \bibinfo {pages} {116402} (\bibinfo {year}
  {2015})}\BibitemShut {NoStop}%
\bibitem [{\citenamefont {Qin}\ \emph {et~al.}(2017)\citenamefont {Qin},
  \citenamefont {Shi},\ and\ \citenamefont {Zhang}}]{Qin:2017}%
  \BibitemOpen
  \bibfield  {author} {\bibinfo {author} {\bibfnamefont {M.}~\bibnamefont
  {Qin}}, \bibinfo {author} {\bibfnamefont {H.}~\bibnamefont {Shi}}, \ and\
  \bibinfo {author} {\bibfnamefont {S.}~\bibnamefont {Zhang}},\ }\href
  {\doibase 10.1103/PhysRevB.96.075156} {\bibfield  {journal} {\bibinfo
  {journal} {Phys. Rev. B}\ }\textbf {\bibinfo {volume} {96}},\ \bibinfo
  {pages} {075156} (\bibinfo {year} {2017})}\BibitemShut {NoStop}%
\bibitem [{\citenamefont {LeBlanc}\ \emph {et~al.}(2019)\citenamefont
  {LeBlanc}, \citenamefont {Li}, \citenamefont {Chen}, \citenamefont {Levy},
  \citenamefont {Antipov}, \citenamefont {Millis},\ and\ \citenamefont
  {Gull}}]{LeBlanc:2019}%
  \BibitemOpen
  \bibfield  {author} {\bibinfo {author} {\bibfnamefont {J.~P.~F.}\
  \bibnamefont {LeBlanc}}, \bibinfo {author} {\bibfnamefont {S.}~\bibnamefont
  {Li}}, \bibinfo {author} {\bibfnamefont {X.}~\bibnamefont {Chen}}, \bibinfo
  {author} {\bibfnamefont {R.}~\bibnamefont {Levy}}, \bibinfo {author}
  {\bibfnamefont {A.~E.}\ \bibnamefont {Antipov}}, \bibinfo {author}
  {\bibfnamefont {A.~J.}\ \bibnamefont {Millis}}, \ and\ \bibinfo {author}
  {\bibfnamefont {E.}~\bibnamefont {Gull}},\ }\href {\doibase
  10.1103/PhysRevB.100.075123} {\bibfield  {journal} {\bibinfo  {journal}
  {Phys. Rev. B}\ }\textbf {\bibinfo {volume} {100}},\ \bibinfo {pages}
  {075123} (\bibinfo {year} {2019})}\BibitemShut {NoStop}%
\bibitem [{\citenamefont {Hille}\ \emph {et~al.}(2020)\citenamefont {Hille},
  \citenamefont {Kugler}, \citenamefont {Eckhardt}, \citenamefont {He},
  \citenamefont {Kauch}, \citenamefont {Honerkamp}, \citenamefont {Toschi},\
  and\ \citenamefont {Andergassen}}]{Hille:2020}%
  \BibitemOpen
  \bibfield  {author} {\bibinfo {author} {\bibfnamefont {C.}~\bibnamefont
  {Hille}}, \bibinfo {author} {\bibfnamefont {F.~B.}\ \bibnamefont {Kugler}},
  \bibinfo {author} {\bibfnamefont {C.~J.}\ \bibnamefont {Eckhardt}}, \bibinfo
  {author} {\bibfnamefont {Y.-Y.}\ \bibnamefont {He}}, \bibinfo {author}
  {\bibfnamefont {A.}~\bibnamefont {Kauch}}, \bibinfo {author} {\bibfnamefont
  {C.}~\bibnamefont {Honerkamp}}, \bibinfo {author} {\bibfnamefont
  {A.}~\bibnamefont {Toschi}}, \ and\ \bibinfo {author} {\bibfnamefont
  {S.}~\bibnamefont {Andergassen}},\ }\href@noop {} {\bibfield  {journal}
  {\bibinfo  {journal} {arXiv:2002.02733}\ } (\bibinfo {year}
  {2020})}\BibitemShut {NoStop}%
\bibitem [{\citenamefont {Bergeron}\ and\ \citenamefont
  {Tremblay}(2016)}]{bergeron:2016}%
  \BibitemOpen
  \bibfield  {author} {\bibinfo {author} {\bibfnamefont {D.}~\bibnamefont
  {Bergeron}}\ and\ \bibinfo {author} {\bibfnamefont {A.-M.~S.}\ \bibnamefont
  {Tremblay}},\ }\href {\doibase 10.1103/PhysRevE.94.023303} {\bibfield
  {journal} {\bibinfo  {journal} {Phys. Rev. E}\ }\textbf {\bibinfo {volume}
  {94}},\ \bibinfo {pages} {023303} (\bibinfo {year} {2016})}\BibitemShut
  {NoStop}%
\bibitem [{\citenamefont {{Levy}}\ \emph {et~al.}(2017)\citenamefont {{Levy}},
  \citenamefont {{LeBlanc}},\ and\ \citenamefont {{Gull}}}]{Levy2016}%
  \BibitemOpen
  \bibfield  {author} {\bibinfo {author} {\bibfnamefont {R.}~\bibnamefont
  {{Levy}}}, \bibinfo {author} {\bibfnamefont {J.~P.~F.}\ \bibnamefont
  {{LeBlanc}}}, \ and\ \bibinfo {author} {\bibfnamefont {E.}~\bibnamefont
  {{Gull}}},\ }\href {\doibase 10.1016/j.cpc.2017.01.018} {\bibfield  {journal}
  {\bibinfo  {journal} {Comp. Phys. Comm.}\ }\textbf {\bibinfo {volume}
  {215}},\ \bibinfo {pages} {149} (\bibinfo {year} {2017})}\BibitemShut
  {NoStop}%
\bibitem [{\citenamefont {Gaenko}\ \emph {et~al.}(2017)\citenamefont {Gaenko},
  \citenamefont {Antipov}, \citenamefont {Carcassi}, \citenamefont {Chen},
  \citenamefont {Chen}, \citenamefont {Dong}, \citenamefont {Gamper},
  \citenamefont {Gukelberger}, \citenamefont {Igarashi}, \citenamefont
  {Iskakov}, \citenamefont {Könz}, \citenamefont {LeBlanc}, \citenamefont
  {Levy}, \citenamefont {Ma}, \citenamefont {Paki}, \citenamefont {Shinaoka},
  \citenamefont {Todo}, \citenamefont {Troyer},\ and\ \citenamefont
  {Gull}}]{Gaenko17}%
  \BibitemOpen
  \bibfield  {author} {\bibinfo {author} {\bibfnamefont {A.}~\bibnamefont
  {Gaenko}}, \bibinfo {author} {\bibfnamefont {A.}~\bibnamefont {Antipov}},
  \bibinfo {author} {\bibfnamefont {G.}~\bibnamefont {Carcassi}}, \bibinfo
  {author} {\bibfnamefont {T.}~\bibnamefont {Chen}}, \bibinfo {author}
  {\bibfnamefont {X.}~\bibnamefont {Chen}}, \bibinfo {author} {\bibfnamefont
  {Q.}~\bibnamefont {Dong}}, \bibinfo {author} {\bibfnamefont {L.}~\bibnamefont
  {Gamper}}, \bibinfo {author} {\bibfnamefont {J.}~\bibnamefont {Gukelberger}},
  \bibinfo {author} {\bibfnamefont {R.}~\bibnamefont {Igarashi}}, \bibinfo
  {author} {\bibfnamefont {S.}~\bibnamefont {Iskakov}}, \bibinfo {author}
  {\bibfnamefont {M.}~\bibnamefont {Könz}}, \bibinfo {author} {\bibfnamefont
  {J.}~\bibnamefont {LeBlanc}}, \bibinfo {author} {\bibfnamefont
  {R.}~\bibnamefont {Levy}}, \bibinfo {author} {\bibfnamefont {P.}~\bibnamefont
  {Ma}}, \bibinfo {author} {\bibfnamefont {J.}~\bibnamefont {Paki}}, \bibinfo
  {author} {\bibfnamefont {H.}~\bibnamefont {Shinaoka}}, \bibinfo {author}
  {\bibfnamefont {S.}~\bibnamefont {Todo}}, \bibinfo {author} {\bibfnamefont
  {M.}~\bibnamefont {Troyer}}, \ and\ \bibinfo {author} {\bibfnamefont
  {E.}~\bibnamefont {Gull}},\ }\href {\doibase
  https://doi.org/10.1016/j.cpc.2016.12.009} {\bibfield  {journal} {\bibinfo
  {journal} {Computer Physics Communications}\ }\textbf {\bibinfo {volume}
  {213}},\ \bibinfo {pages} {235 } (\bibinfo {year} {2017})}\BibitemShut
  {NoStop}%
\bibitem [{\citenamefont {Gaenko}\ \emph {et~al.}(2016)\citenamefont {Gaenko},
  \citenamefont {Gull}, \citenamefont {Antipov}, \citenamefont {Gamper},
  \citenamefont {Carcassi}, \citenamefont {Paki}, \citenamefont {Levy},
  \citenamefont {Dolfi}, \citenamefont {Greitemann},\ and\ \citenamefont
  {LeBlanc}}]{ALPSCore}%
  \BibitemOpen
  \bibfield  {author} {\bibinfo {author} {\bibfnamefont {A.}~\bibnamefont
  {Gaenko}}, \bibinfo {author} {\bibfnamefont {E.}~\bibnamefont {Gull}},
  \bibinfo {author} {\bibfnamefont {A.~E.}\ \bibnamefont {Antipov}}, \bibinfo
  {author} {\bibfnamefont {L.}~\bibnamefont {Gamper}}, \bibinfo {author}
  {\bibfnamefont {G.}~\bibnamefont {Carcassi}}, \bibinfo {author}
  {\bibfnamefont {J.}~\bibnamefont {Paki}}, \bibinfo {author} {\bibfnamefont
  {R.}~\bibnamefont {Levy}}, \bibinfo {author} {\bibfnamefont {M.}~\bibnamefont
  {Dolfi}}, \bibinfo {author} {\bibfnamefont {J.}~\bibnamefont {Greitemann}}, \
  and\ \bibinfo {author} {\bibfnamefont {J.~P.}\ \bibnamefont {LeBlanc}},\
  }\href {\doibase 10.5281/zenodo.50203} {\enquote {\bibinfo {title} {Alpscore:
  Version 0.5.4},}\ } (\bibinfo {year} {2016})\BibitemShut {NoStop}%
\bibitem [{\citenamefont {Wallerberger}\ \emph {et~al.}(2018)\citenamefont
  {Wallerberger}, \citenamefont {Iskakov}, \citenamefont {Gaenko},
  \citenamefont {Kleinhenz}, \citenamefont {Krivenko}, \citenamefont {Levy},
  \citenamefont {Li}, \citenamefont {Shinaoka}, \citenamefont {Todo},
  \citenamefont {Chen}, \citenamefont {Chen}, \citenamefont {LeBlanc},
  \citenamefont {Paki}, \citenamefont {Terletska}, \citenamefont {Troyer},\
  and\ \citenamefont {Gull}}]{alpscore_v2}%
  \BibitemOpen
  \bibfield  {author} {\bibinfo {author} {\bibfnamefont {M.}~\bibnamefont
  {Wallerberger}}, \bibinfo {author} {\bibfnamefont {S.}~\bibnamefont
  {Iskakov}}, \bibinfo {author} {\bibfnamefont {A.}~\bibnamefont {Gaenko}},
  \bibinfo {author} {\bibfnamefont {J.}~\bibnamefont {Kleinhenz}}, \bibinfo
  {author} {\bibfnamefont {I.}~\bibnamefont {Krivenko}}, \bibinfo {author}
  {\bibfnamefont {R.}~\bibnamefont {Levy}}, \bibinfo {author} {\bibfnamefont
  {J.}~\bibnamefont {Li}}, \bibinfo {author} {\bibfnamefont {H.}~\bibnamefont
  {Shinaoka}}, \bibinfo {author} {\bibfnamefont {S.}~\bibnamefont {Todo}},
  \bibinfo {author} {\bibfnamefont {T.}~\bibnamefont {Chen}}, \bibinfo {author}
  {\bibfnamefont {X.}~\bibnamefont {Chen}}, \bibinfo {author} {\bibfnamefont
  {J.~P.~F.}\ \bibnamefont {LeBlanc}}, \bibinfo {author} {\bibfnamefont
  {J.~E.}\ \bibnamefont {Paki}}, \bibinfo {author} {\bibfnamefont
  {H.}~\bibnamefont {Terletska}}, \bibinfo {author} {\bibfnamefont
  {M.}~\bibnamefont {Troyer}}, \ and\ \bibinfo {author} {\bibfnamefont
  {E.}~\bibnamefont {Gull}},\ }\href@noop {} {\bibfield  {journal} {\bibinfo
  {journal} {arXiv:1811.08331}\ } (\bibinfo {year} {2018})}\BibitemShut
  {NoStop}%
\bibitem [{\citenamefont {Vu\ifmmode \check{c}\else \v{c}\fi{}i\ifmmode
  \check{c}\else \v{c}\fi{}evi\ifmmode~\acute{c}\else \'{c}\fi{}}\ \emph
  {et~al.}(2019)\citenamefont {Vu\ifmmode \check{c}\else \v{c}\fi{}i\ifmmode
  \check{c}\else \v{c}\fi{}evi\ifmmode~\acute{c}\else \'{c}\fi{}},
  \citenamefont {Kokalj}, \citenamefont {\ifmmode~\check{Z}\else
  \v{Z}\fi{}itko}, \citenamefont {Wentzell}, \citenamefont
  {Tanaskovi\ifmmode~\acute{c}\else \'{c}\fi{}},\ and\ \citenamefont
  {Mravlje}}]{Mravlje}%
  \BibitemOpen
  \bibfield  {author} {\bibinfo {author} {\bibfnamefont {J.}~\bibnamefont
  {Vu\ifmmode \check{c}\else \v{c}\fi{}i\ifmmode \check{c}\else
  \v{c}\fi{}evi\ifmmode~\acute{c}\else \'{c}\fi{}}}, \bibinfo {author}
  {\bibfnamefont {J.}~\bibnamefont {Kokalj}}, \bibinfo {author} {\bibfnamefont
  {R.}~\bibnamefont {\ifmmode~\check{Z}\else \v{Z}\fi{}itko}}, \bibinfo
  {author} {\bibfnamefont {N.}~\bibnamefont {Wentzell}}, \bibinfo {author}
  {\bibfnamefont {D.}~\bibnamefont {Tanaskovi\ifmmode~\acute{c}\else
  \'{c}\fi{}}}, \ and\ \bibinfo {author} {\bibfnamefont {J.}~\bibnamefont
  {Mravlje}},\ }\href {\doibase 10.1103/PhysRevLett.123.036601} {\bibfield
  {journal} {\bibinfo  {journal} {Phys. Rev. Lett.}\ }\textbf {\bibinfo
  {volume} {123}},\ \bibinfo {pages} {036601} (\bibinfo {year}
  {2019})}\BibitemShut {NoStop}%
\bibitem [{\citenamefont {Huang}\ \emph {et~al.}(2019)\citenamefont {Huang},
  \citenamefont {Sheppard}, \citenamefont {Moritz},\ and\ \citenamefont
  {Devereaux}}]{Huang2019}%
  \BibitemOpen
  \bibfield  {author} {\bibinfo {author} {\bibfnamefont {E.~W.}\ \bibnamefont
  {Huang}}, \bibinfo {author} {\bibfnamefont {R.}~\bibnamefont {Sheppard}},
  \bibinfo {author} {\bibfnamefont {B.}~\bibnamefont {Moritz}}, \ and\ \bibinfo
  {author} {\bibfnamefont {T.~P.}\ \bibnamefont {Devereaux}},\ }\href {\doibase
  10.1126/science.aau7063} {\bibfield  {journal} {\bibinfo  {journal}
  {Science}\ }\textbf {\bibinfo {volume} {366}},\ \bibinfo {pages} {987}
  (\bibinfo {year} {2019})}\BibitemShut {NoStop}%
\bibitem [{\citenamefont {Vu\ifmmode \check{c}\else \v{c}\fi{}i\ifmmode
  \check{c}\else \v{c}\fi{}evi\ifmmode~\acute{c}\else \'{c}\fi{}}\ and\
  \citenamefont {Ferrero}(2020)}]{Ferrero:AMI}%
  \BibitemOpen
  \bibfield  {author} {\bibinfo {author} {\bibfnamefont {J.}~\bibnamefont
  {Vu\ifmmode \check{c}\else \v{c}\fi{}i\ifmmode \check{c}\else
  \v{c}\fi{}evi\ifmmode~\acute{c}\else \'{c}\fi{}}}\ and\ \bibinfo {author}
  {\bibfnamefont {M.}~\bibnamefont {Ferrero}},\ }\href {\doibase
  10.1103/PhysRevB.101.075113} {\bibfield  {journal} {\bibinfo  {journal}
  {Phys. Rev. B}\ }\textbf {\bibinfo {volume} {101}},\ \bibinfo {pages}
  {075113} (\bibinfo {year} {2020})}\BibitemShut {NoStop}%
\bibitem [{\citenamefont {Taheridehkordi}\ \emph {et~al.}(2019)\citenamefont
  {Taheridehkordi}, \citenamefont {Curnoe},\ and\ \citenamefont
  {LeBlanc}}]{AMI}%
  \BibitemOpen
  \bibfield  {author} {\bibinfo {author} {\bibfnamefont {A.}~\bibnamefont
  {Taheridehkordi}}, \bibinfo {author} {\bibfnamefont {S.~H.}\ \bibnamefont
  {Curnoe}}, \ and\ \bibinfo {author} {\bibfnamefont {J.~P.~F.}\ \bibnamefont
  {LeBlanc}},\ }\href {\doibase 10.1103/PhysRevB.99.035120} {\bibfield
  {journal} {\bibinfo  {journal} {Phys. Rev. B}\ }\textbf {\bibinfo {volume}
  {99}},\ \bibinfo {pages} {035120} (\bibinfo {year} {2019})}\BibitemShut
  {NoStop}%
\bibitem [{\citenamefont {Loh}\ \emph {et~al.}(1990)\citenamefont {Loh},
  \citenamefont {Gubernatis}, \citenamefont {Scalettar}, \citenamefont {White},
  \citenamefont {Scalapino},\ and\ \citenamefont {Sugar}}]{Loh}%
  \BibitemOpen
  \bibfield  {author} {\bibinfo {author} {\bibfnamefont {E.~Y.}\ \bibnamefont
  {Loh}}, \bibinfo {author} {\bibfnamefont {J.~E.}\ \bibnamefont {Gubernatis}},
  \bibinfo {author} {\bibfnamefont {R.~T.}\ \bibnamefont {Scalettar}}, \bibinfo
  {author} {\bibfnamefont {S.~R.}\ \bibnamefont {White}}, \bibinfo {author}
  {\bibfnamefont {D.~J.}\ \bibnamefont {Scalapino}}, \ and\ \bibinfo {author}
  {\bibfnamefont {R.~L.}\ \bibnamefont {Sugar}},\ }\href {\doibase
  10.1103/PhysRevB.41.9301} {\bibfield  {journal} {\bibinfo  {journal} {Phys.
  Rev. B}\ }\textbf {\bibinfo {volume} {41}},\ \bibinfo {pages} {9301}
  (\bibinfo {year} {1990})}\BibitemShut {NoStop}%
\bibitem [{\citenamefont {Chandrasekharan}\ and\ \citenamefont
  {Wiese}(1999)}]{Chandrasekharan}%
  \BibitemOpen
  \bibfield  {author} {\bibinfo {author} {\bibfnamefont {S.}~\bibnamefont
  {Chandrasekharan}}\ and\ \bibinfo {author} {\bibfnamefont {U.-J.}\
  \bibnamefont {Wiese}},\ }\href {\doibase 10.1103/PhysRevLett.83.3116}
  {\bibfield  {journal} {\bibinfo  {journal} {Phys. Rev. Lett.}\ }\textbf
  {\bibinfo {volume} {83}},\ \bibinfo {pages} {3116} (\bibinfo {year}
  {1999})}\BibitemShut {NoStop}%
\bibitem [{\citenamefont {Taheridehkordi}\ \emph {et~al.}(2020)\citenamefont
  {Taheridehkordi}, \citenamefont {Curnoe},\ and\ \citenamefont
  {LeBlanc}}]{GIT}%
  \BibitemOpen
  \bibfield  {author} {\bibinfo {author} {\bibfnamefont {A.}~\bibnamefont
  {Taheridehkordi}}, \bibinfo {author} {\bibfnamefont {S.~H.}\ \bibnamefont
  {Curnoe}}, \ and\ \bibinfo {author} {\bibfnamefont {J.~P.~F.}\ \bibnamefont
  {LeBlanc}},\ }\href {\doibase 10.1103/PhysRevB.101.125109} {\bibfield
  {journal} {\bibinfo  {journal} {Phys. Rev. B}\ }\textbf {\bibinfo {volume}
  {101}},\ \bibinfo {pages} {125109} (\bibinfo {year} {2020})}\BibitemShut
  {NoStop}%
\bibitem [{\citenamefont {Bohm}\ and\ \citenamefont
  {Pines}(1951)}]{Bohm_RPA_1}%
  \BibitemOpen
  \bibfield  {author} {\bibinfo {author} {\bibfnamefont {D.}~\bibnamefont
  {Bohm}}\ and\ \bibinfo {author} {\bibfnamefont {D.}~\bibnamefont {Pines}},\
  }\href {\doibase 10.1103/PhysRev.82.625} {\bibfield  {journal} {\bibinfo
  {journal} {Phys. Rev.}\ }\textbf {\bibinfo {volume} {82}},\ \bibinfo {pages}
  {625} (\bibinfo {year} {1951})}\BibitemShut {NoStop}%
\bibitem [{\citenamefont {Bohm}\ and\ \citenamefont
  {Pines}(1953)}]{Bohm_RPA_2}%
  \BibitemOpen
  \bibfield  {author} {\bibinfo {author} {\bibfnamefont {D.}~\bibnamefont
  {Bohm}}\ and\ \bibinfo {author} {\bibfnamefont {D.}~\bibnamefont {Pines}},\
  }\href {\doibase 10.1103/PhysRev.92.609} {\bibfield  {journal} {\bibinfo
  {journal} {Phys. Rev.}\ }\textbf {\bibinfo {volume} {92}},\ \bibinfo {pages}
  {609} (\bibinfo {year} {1953})}\BibitemShut {NoStop}%
\bibitem [{\citenamefont {Fukuyama}\ and\ \citenamefont
  {Hasegawa}(1990)}]{Fukuyama:1990}%
  \BibitemOpen
  \bibfield  {author} {\bibinfo {author} {\bibfnamefont {H.}~\bibnamefont
  {Fukuyama}}\ and\ \bibinfo {author} {\bibfnamefont {Y.}~\bibnamefont
  {Hasegawa}},\ }\href {\doibase 10.1143/PTP.101.441} {\bibfield  {journal}
  {\bibinfo  {journal} {Progress of Theoretical Physics Supplement}\ }\textbf
  {\bibinfo {volume} {101}},\ \bibinfo {pages} {441} (\bibinfo {year}
  {1990})}\BibitemShut {NoStop}%
\bibitem [{\citenamefont {Gukelberger}\ \emph {et~al.}(2015)\citenamefont
  {Gukelberger}, \citenamefont {Huang},\ and\ \citenamefont
  {Werner}}]{Gukelberger:2015}%
  \BibitemOpen
  \bibfield  {author} {\bibinfo {author} {\bibfnamefont {J.}~\bibnamefont
  {Gukelberger}}, \bibinfo {author} {\bibfnamefont {L.}~\bibnamefont {Huang}},
  \ and\ \bibinfo {author} {\bibfnamefont {P.}~\bibnamefont {Werner}},\ }\href
  {\doibase 10.1103/PhysRevB.91.235114} {\bibfield  {journal} {\bibinfo
  {journal} {Phys. Rev. B}\ }\textbf {\bibinfo {volume} {91}},\ \bibinfo
  {pages} {235114} (\bibinfo {year} {2015})}\BibitemShut {NoStop}%
\bibitem [{\citenamefont {Yoshimi}\ \emph {et~al.}(2009)\citenamefont
  {Yoshimi}, \citenamefont {Kato},\ and\ \citenamefont
  {Maebashi}}]{Yoshimi:2009}%
  \BibitemOpen
  \bibfield  {author} {\bibinfo {author} {\bibfnamefont {K.}~\bibnamefont
  {Yoshimi}}, \bibinfo {author} {\bibfnamefont {T.}~\bibnamefont {Kato}}, \
  and\ \bibinfo {author} {\bibfnamefont {H.}~\bibnamefont {Maebashi}},\ }\href
  {\doibase 10.1143/JPSJ.78.104002} {\bibfield  {journal} {\bibinfo  {journal}
  {Journal of the Physical Society of Japan}\ }\textbf {\bibinfo {volume}
  {78}},\ \bibinfo {pages} {104002} (\bibinfo {year} {2009})}\BibitemShut
  {NoStop}%
\bibitem [{\citenamefont {Lieb}\ and\ \citenamefont {Wu}(1968)}]{Lieb:1968}%
  \BibitemOpen
  \bibfield  {author} {\bibinfo {author} {\bibfnamefont {E.~H.}\ \bibnamefont
  {Lieb}}\ and\ \bibinfo {author} {\bibfnamefont {F.~Y.}\ \bibnamefont {Wu}},\
  }\href {\doibase 10.1103/PhysRevLett.21.192.2} {\bibfield  {journal}
  {\bibinfo  {journal} {Phys. Rev. Lett.}\ }\textbf {\bibinfo {volume} {21}},\
  \bibinfo {pages} {192} (\bibinfo {year} {1968})}\BibitemShut {NoStop}%
\bibitem [{\citenamefont {Vicente~Alvarez}\ \emph {et~al.}(1996)\citenamefont
  {Vicente~Alvarez}, \citenamefont {Balseiro},\ and\ \citenamefont
  {Ceccatto}}]{Alvarez:1996}%
  \BibitemOpen
  \bibfield  {author} {\bibinfo {author} {\bibfnamefont {J.~J.}\ \bibnamefont
  {Vicente~Alvarez}}, \bibinfo {author} {\bibfnamefont {C.~A.}\ \bibnamefont
  {Balseiro}}, \ and\ \bibinfo {author} {\bibfnamefont {H.~A.}\ \bibnamefont
  {Ceccatto}},\ }\href {\doibase 10.1103/PhysRevB.54.11207} {\bibfield
  {journal} {\bibinfo  {journal} {Phys. Rev. B}\ }\textbf {\bibinfo {volume}
  {54}},\ \bibinfo {pages} {11207} (\bibinfo {year} {1996})}\BibitemShut
  {NoStop}%
\bibitem [{\citenamefont {Masumizu}\ and\ \citenamefont
  {Sogo}(2005)}]{Masumizu:2005}%
  \BibitemOpen
  \bibfield  {author} {\bibinfo {author} {\bibfnamefont {A.}~\bibnamefont
  {Masumizu}}\ and\ \bibinfo {author} {\bibfnamefont {K.}~\bibnamefont
  {Sogo}},\ }\href {\doibase 10.1103/PhysRevB.72.115107} {\bibfield  {journal}
  {\bibinfo  {journal} {Phys. Rev. B}\ }\textbf {\bibinfo {volume} {72}},\
  \bibinfo {pages} {115107} (\bibinfo {year} {2005})}\BibitemShut {NoStop}%
\bibitem [{\citenamefont {Feynman}(1949)}]{Feynman}%
  \BibitemOpen
  \bibfield  {author} {\bibinfo {author} {\bibfnamefont {R.~P.}\ \bibnamefont
  {Feynman}},\ }\href {\doibase 10.1103/PhysRev.76.769} {\bibfield  {journal}
  {\bibinfo  {journal} {Phys. Rev.}\ }\textbf {\bibinfo {volume} {76}},\
  \bibinfo {pages} {769} (\bibinfo {year} {1949})}\BibitemShut {NoStop}%
\bibitem [{\citenamefont {Baym}\ and\ \citenamefont {Kadanoff}(1961)}]{Baym}%
  \BibitemOpen
  \bibfield  {author} {\bibinfo {author} {\bibfnamefont {G.}~\bibnamefont
  {Baym}}\ and\ \bibinfo {author} {\bibfnamefont {L.~P.}\ \bibnamefont
  {Kadanoff}},\ }\href {\doibase 10.1103/PhysRev.124.287} {\bibfield  {journal}
  {\bibinfo  {journal} {Phys. Rev.}\ }\textbf {\bibinfo {volume} {124}},\
  \bibinfo {pages} {287} (\bibinfo {year} {1961})}\BibitemShut {NoStop}%
\bibitem [{\citenamefont {Luttinger}\ and\ \citenamefont
  {Ward}(1960)}]{Luttinger:1960}%
  \BibitemOpen
  \bibfield  {author} {\bibinfo {author} {\bibfnamefont {J.~M.}\ \bibnamefont
  {Luttinger}}\ and\ \bibinfo {author} {\bibfnamefont {J.~C.}\ \bibnamefont
  {Ward}},\ }\href {\doibase 10.1103/PhysRev.118.1417} {\bibfield  {journal}
  {\bibinfo  {journal} {Phys. Rev.}\ }\textbf {\bibinfo {volume} {118}},\
  \bibinfo {pages} {1417} (\bibinfo {year} {1960})}\BibitemShut {NoStop}%
\bibitem [{\citenamefont {Zlati\ifmmode~\acute{c}\else \'{c}\fi{}}\ \emph
  {et~al.}(2000)\citenamefont {Zlati\ifmmode~\acute{c}\else \'{c}\fi{}},
  \citenamefont {Horvati\ifmmode~\acute{c}\else \'{c}\fi{}}, \citenamefont
  {Doli\ifmmode~\check{c}\else \v{c}\fi{}ki}, \citenamefont {Grabowski},
  \citenamefont {Entel},\ and\ \citenamefont {Schotte}}]{Zlatic}%
  \BibitemOpen
  \bibfield  {author} {\bibinfo {author} {\bibfnamefont {V.}~\bibnamefont
  {Zlati\ifmmode~\acute{c}\else \'{c}\fi{}}}, \bibinfo {author} {\bibfnamefont
  {B.}~\bibnamefont {Horvati\ifmmode~\acute{c}\else \'{c}\fi{}}}, \bibinfo
  {author} {\bibfnamefont {B.}~\bibnamefont {Doli\ifmmode~\check{c}\else
  \v{c}\fi{}ki}}, \bibinfo {author} {\bibfnamefont {S.}~\bibnamefont
  {Grabowski}}, \bibinfo {author} {\bibfnamefont {P.}~\bibnamefont {Entel}}, \
  and\ \bibinfo {author} {\bibfnamefont {K.-D.}\ \bibnamefont {Schotte}},\
  }\href {\doibase 10.1103/PhysRevB.63.035104} {\bibfield  {journal} {\bibinfo
  {journal} {Phys. Rev. B}\ }\textbf {\bibinfo {volume} {63}},\ \bibinfo
  {pages} {035104} (\bibinfo {year} {2000})}\BibitemShut {NoStop}%
\bibitem [{\citenamefont {Daul}\ and\ \citenamefont
  {Dzierzawa}(1997)}]{Daul1997}%
  \BibitemOpen
  \bibfield  {author} {\bibinfo {author} {\bibfnamefont {S.}~\bibnamefont
  {Daul}}\ and\ \bibinfo {author} {\bibfnamefont {M.}~\bibnamefont
  {Dzierzawa}},\ }\href {\doibase 10.1007/s002570050332} {\bibfield  {journal}
  {\bibinfo  {journal} {Zeitschrift f{\"u}r Physik B Condensed Matter}\
  }\textbf {\bibinfo {volume} {103}},\ \bibinfo {pages} {41} (\bibinfo {year}
  {1997})}\BibitemShut {NoStop}%
\bibitem [{\citenamefont {Tan}(2006)}]{Zhiqiang}%
  \BibitemOpen
  \bibfield  {author} {\bibinfo {author} {\bibfnamefont {Z.}~\bibnamefont
  {Tan}},\ }\href@noop {} {\bibfield  {journal} {\bibinfo  {journal} {Journal
  of Computational and Graphical Statistics}\ }\textbf {\bibinfo {volume} {15}}
  (\bibinfo {year} {2006})}\BibitemShut {NoStop}%
\bibitem [{\citenamefont {Hahn}(2005)}]{CUBA}%
  \BibitemOpen
  \bibfield  {author} {\bibinfo {author} {\bibfnamefont {T.}~\bibnamefont
  {Hahn}},\ }\href {\doibase https://doi.org/10.1016/j.cpc.2005.01.010}
  {\bibfield  {journal} {\bibinfo  {journal} {Computer Physics Communications}\
  }\textbf {\bibinfo {volume} {168}},\ \bibinfo {pages} {78 } (\bibinfo {year}
  {2005})}\BibitemShut {NoStop}%
\bibitem [{\citenamefont {Kugler}(2018)}]{Kugler}%
  \BibitemOpen
  \bibfield  {author} {\bibinfo {author} {\bibfnamefont {F.~B.}\ \bibnamefont
  {Kugler}},\ }\href {\doibase 10.1103/PhysRevE.98.023303} {\bibfield
  {journal} {\bibinfo  {journal} {Phys. Rev. E}\ }\textbf {\bibinfo {volume}
  {98}},\ \bibinfo {pages} {023303} (\bibinfo {year} {2018})}\BibitemShut
  {NoStop}%
\bibitem [{\citenamefont {Tarantino}\ \emph {et~al.}(2017)\citenamefont
  {Tarantino}, \citenamefont {Romaniello}, \citenamefont {Berger},\ and\
  \citenamefont {Reining}}]{tarantino:2017}%
  \BibitemOpen
  \bibfield  {author} {\bibinfo {author} {\bibfnamefont {W.}~\bibnamefont
  {Tarantino}}, \bibinfo {author} {\bibfnamefont {P.}~\bibnamefont
  {Romaniello}}, \bibinfo {author} {\bibfnamefont {J.~A.}\ \bibnamefont
  {Berger}}, \ and\ \bibinfo {author} {\bibfnamefont {L.}~\bibnamefont
  {Reining}},\ }\href {\doibase 10.1103/PhysRevB.96.045124} {\bibfield
  {journal} {\bibinfo  {journal} {Phys. Rev. B}\ }\textbf {\bibinfo {volume}
  {96}},\ \bibinfo {pages} {045124} (\bibinfo {year} {2017})}\BibitemShut
  {NoStop}%
\bibitem [{\citenamefont {{Fetter}}\ and\ \citenamefont
  {{Walecka}}(2003)}]{Fetter}%
  \BibitemOpen
  \bibfield  {author} {\bibinfo {author} {\bibfnamefont {A.}~\bibnamefont
  {{Fetter}}}\ and\ \bibinfo {author} {\bibfnamefont {J.}~\bibnamefont
  {{Walecka}}},\ }\href@noop {} {\emph {\bibinfo {title} {Quantum Theory of
  Many-particle Systems}}}\ (\bibinfo  {publisher} {Dover Publications},\
  \bibinfo {address} {Mineola, New York},\ \bibinfo {year} {2003})\BibitemShut
  {NoStop}%
\bibitem [{\citenamefont {Arzhang}\ \emph {et~al.}(2020)\citenamefont
  {Arzhang}, \citenamefont {Antipov},\ and\ \citenamefont
  {LeBlanc}}]{behnam:2020}%
  \BibitemOpen
  \bibfield  {author} {\bibinfo {author} {\bibfnamefont {B.}~\bibnamefont
  {Arzhang}}, \bibinfo {author} {\bibfnamefont {A.~E.}\ \bibnamefont
  {Antipov}}, \ and\ \bibinfo {author} {\bibfnamefont {J.~P.~F.}\ \bibnamefont
  {LeBlanc}},\ }\href {\doibase 10.1103/PhysRevB.101.014430} {\bibfield
  {journal} {\bibinfo  {journal} {Phys. Rev. B}\ }\textbf {\bibinfo {volume}
  {101}},\ \bibinfo {pages} {014430} (\bibinfo {year} {2020})}\BibitemShut
  {NoStop}%
\bibitem [{\citenamefont {Antipov}\ \emph {et~al.}(2015)\citenamefont
  {Antipov}, \citenamefont {LeBlanc},\ and\ \citenamefont {Gull}}]{opendf}%
  \BibitemOpen
  \bibfield  {author} {\bibinfo {author} {\bibfnamefont {A.~E.}\ \bibnamefont
  {Antipov}}, \bibinfo {author} {\bibfnamefont {J.~P.~F.}\ \bibnamefont
  {LeBlanc}}, \ and\ \bibinfo {author} {\bibfnamefont {E.}~\bibnamefont
  {Gull}},\ }\href {\doibase http://dx.doi.org/10.1016/j.phpro.2015.07.107}
  {\bibfield  {journal} {\bibinfo  {journal} {Physics Procedia}\ }\textbf
  {\bibinfo {volume} {68}},\ \bibinfo {pages} {43 } (\bibinfo {year} {2015})},\
  \bibinfo {note} {proceedings of the 28th Workshop on Computer Simulation
  Studies in Condensed Matter Physics (CSP2015)}\BibitemShut {NoStop}%
\bibitem [{\citenamefont {Gukelberger}\ \emph {et~al.}(2017)\citenamefont
  {Gukelberger}, \citenamefont {Kozik},\ and\ \citenamefont
  {Hafermann}}]{Gukelberger:DF}%
  \BibitemOpen
  \bibfield  {author} {\bibinfo {author} {\bibfnamefont {J.}~\bibnamefont
  {Gukelberger}}, \bibinfo {author} {\bibfnamefont {E.}~\bibnamefont {Kozik}},
  \ and\ \bibinfo {author} {\bibfnamefont {H.}~\bibnamefont {Hafermann}},\
  }\href {\doibase 10.1103/PhysRevB.96.035152} {\bibfield  {journal} {\bibinfo
  {journal} {Phys. Rev. B}\ }\textbf {\bibinfo {volume} {96}},\ \bibinfo
  {pages} {035152} (\bibinfo {year} {2017})}\BibitemShut {NoStop}%
\bibitem [{\citenamefont {Hayden}\ \emph {et~al.}(1990)\citenamefont {Hayden},
  \citenamefont {Aeppli}, \citenamefont {Mook}, \citenamefont {Cheong},\ and\
  \citenamefont {Fisk}}]{hayden:1990}%
  \BibitemOpen
  \bibfield  {author} {\bibinfo {author} {\bibfnamefont {S.~M.}\ \bibnamefont
  {Hayden}}, \bibinfo {author} {\bibfnamefont {G.}~\bibnamefont {Aeppli}},
  \bibinfo {author} {\bibfnamefont {H.~A.}\ \bibnamefont {Mook}}, \bibinfo
  {author} {\bibfnamefont {S.-W.}\ \bibnamefont {Cheong}}, \ and\ \bibinfo
  {author} {\bibfnamefont {Z.}~\bibnamefont {Fisk}},\ }\href {\doibase
  10.1103/PhysRevB.42.10220} {\bibfield  {journal} {\bibinfo  {journal} {Phys.
  Rev. B}\ }\textbf {\bibinfo {volume} {42}},\ \bibinfo {pages} {10220}
  (\bibinfo {year} {1990})}\BibitemShut {NoStop}%
\bibitem [{\citenamefont {Headings}\ \emph {et~al.}(2010)\citenamefont
  {Headings}, \citenamefont {Hayden}, \citenamefont {Coldea},\ and\
  \citenamefont {Perring}}]{headings:2010}%
  \BibitemOpen
  \bibfield  {author} {\bibinfo {author} {\bibfnamefont {N.~S.}\ \bibnamefont
  {Headings}}, \bibinfo {author} {\bibfnamefont {S.~M.}\ \bibnamefont
  {Hayden}}, \bibinfo {author} {\bibfnamefont {R.}~\bibnamefont {Coldea}}, \
  and\ \bibinfo {author} {\bibfnamefont {T.~G.}\ \bibnamefont {Perring}},\
  }\href {\doibase 10.1103/PhysRevLett.105.247001} {\bibfield  {journal}
  {\bibinfo  {journal} {Phys. Rev. Lett.}\ }\textbf {\bibinfo {volume} {105}},\
  \bibinfo {pages} {247001} (\bibinfo {year} {2010})}\BibitemShut {NoStop}%
\end{thebibliography}%

\end{document}